\documentclass[a4paper,12pt]{article}

\usepackage[titletoc,title]{appendix}
\usepackage{amsmath}
\usepackage{amssymb}
\usepackage{slashed}
\usepackage{cancel}

\renewcommand{\d}{\mathrm{d}}
\renewcommand{\t}[1]{\text{#1}}

\numberwithin{equation}{section}
\newcommand{\bea}{\begin{eqnarray}}
\newcommand{\eea}{\end{eqnarray}}

\begin{document}

\begin{titlepage}
\begin{center}

\vspace*{-1.0cm}

\hfill  DMUS-MP-22-06
\\
\vspace{2.0cm}

\renewcommand{\thefootnote}{\fnsymbol{footnote}}
{\Large{\bf Supersymmetric $\t{dS}_4$ solutions in $D=11$ supergravity}}
\vskip1cm
\vskip 1.3cm
M. Di Gioia and J. Gutowski
\vskip 1cm
{\small{\it
Department of Mathematics,
University of Surrey \\
Guildford, GU2 7XH, UK.}\\
\texttt{m.digioia@surrey.ac.uk, j.gutowski@surrey.ac.uk}}

\end{center}
\bigskip
\begin{center}
{\bf Abstract}
\end{center}
Supersymmetric warped product $\t{dS}_4$ solutions in $D=11$ supergravity are classified. 
The Killing spinor is associated with two possible stabilizer groups, $SU(3)$ and $G_2$. We show that there are no solutions to the Killing Spinor equations in the $G_2$ stabilizer case. For the $SU(3)$ stablilzer case, all of the conditions imposed from supersymmetry on the 4-form flux, and the geometry of the internal manifold, are determined in terms of $SU(3)$ invariant spinor bilinears.

\end{titlepage}

\section{Introduction}

De Sitter geometry is of particular interest in terms of string cosmology and also in the context of the holographic principle. De Sitter spacetime plays a central role in the understanding of our present universe. From the work of \cite{Schmidt_1998,Riess_1998,Perlmutter_1999} it has been observed that our 
universe is asymptotically $\t{dS}_4$, corresponding to a very small positive cosmological constant. However, the observed value of the cosmological constant differs by 
many orders of magnitude from the vacuum energy density value predicted by quantum field theory \cite{Weinberg:1988cp, Linde:1974at}. Moreover, in the context of string cosmology there are also difficulties in obtaining de Sitter space via compactification from higher dimensions. In particular, there are no go-theorems proving that smooth warped de Sitter solutions with compact, without boundary, internal manifold cannot be found in ten- and eleven-dimensional supergravity \cite{Gibbons1,Smit,Maldacena1}. Issues relating
to quantum gravity in de Sitter space have been investigated in \cite{witten1}.

In terms of holography, the $\t{AdS}/\t{CFT}$ correspondence relates string theory in Anti-de Sitter (AdS) space to conformal field theories (CFT) defined on an appropriate boundary~\cite{Maldacena_1999}. This has been particularly useful in developing
a deeper understanding of the the microscopic nature of the entropy-area law~\cite{PhysRevD.7.2333,cmp/1103899181}.
In spite of the considerable insights produced via the holographic principle, 
there are still many open issues in this area. Building from the $\t{AdS}_3/\t{CFT}_2$ correspondence proposed by Brown and Henneaux in~\cite{cite-key5545346546256}, 
 the relation between quantum gravity on de Sitter space and conformal field theory on a sphere, the so-called dS/CFT correspondence, was considered in \cite{Maldacena12,Strominger_20012,Strominger_2001}. However, our understanding of
the conjectured dS/CFT correspondence is less complete than for the case of AdS/CFT for a number of reasons. Firstly, in contrast to $\t{AdS}$, there is a lack of de Sitter space solutions in string theory (or in any quantum gravity theory) in which the conjecture can be tested. Also, there are subtle issues with defining the dual CFT on the past and future spheres ${\mathcal{I}}^\pm$, relating to the causal structure
of $\t{dS}$ space. Nevertheless, the macroscopic entropy-area law applies to a very wide class of black holes, including asymptotically flat, asymptotically $\t{AdS}$, and also asymptotically $\t{dS}$ cases. The universality of this law provides strong motivation for understanding de Sitter holography.

Motivated by this, it is of particular interest to systematically understand the different types of de Sitter solutions which are possible in D=10 and D=11 supergravity. Such a classification may provide interesting new applications of the dS/CFT correspondence. As it is possible to embed $\t{dS}_n$ inside both $\mathbb{R}^{1,n}$
and $\t{AdS}_{n+1}$ as a warped product geometry \cite{Gran:2016zxk}, it follows
that the maximally supersymmetric $\t{AdS}_7\times \t{S}^4$ solution, as well as 
$\mathbb{R}^{1,10}$, can both be regarded as examples of warped product $\t{dS}_4$ geometries. However, as we shall establish here, there is a much larger class of 
supersymmetric warped product $\t{dS}_4$ solutions in $D=11$ supergravity than these two very special solutions, and this is also somewhat in contrast to the results of recent analysis of supersymmetric warped product $\t{dS}_n$ geometries for $5 \leq n \leq 10$.

In terms of $D=11$ supergravity, there has been recent progress in the classification of supersymmetric warped product $\t{dS}_n$ geometries for $5 \leq n \leq 10$ \cite{Farotti:2022xsd}. There are a number of different possibilities:

\begin{itemize}

\item For $7 \leq n \leq 10$, the geometry is the maximally supersymmetic $\mathbb{R}^{1,10}$ solution with vanishing 4-form flux.

\item For warped product $\t{dS}_6$ solutions, the solution is either the maximally supersymmetric $\t{AdS}_7 \times {\t{S}}^4$ solution, or $\mathbb{R}^{1,6} \times N$ where
$N$ is a hyper-K\"ahler 4-manifold.

\item The warped product $\t{dS}_5$ solutions are all examples of generalized M5-brane solutions for which the transverse space is $\mathbb{R} \times N$, where $N$ is a hyper-K\"ahler 4-manifold.

\end{itemize}

It is clear from this list that the possible  warped product $\t{dS}_n$ geometries for $5 \leq n \leq 10$ is very highly constrained. In addition, a similar recent analysis of warped product $\t{dS}_n$ solutions in heterotic supergravity \cite{Farotti:2022twf},
including first order $\alpha'$ corrections, has also produced a rather restricted class of such solutions. In this case, for $n \geq 3$, the geometry is $\mathbb{R}^{1,n} \times M_{9-n}$, where $M_{9-n}$ is a $(9-n)$-dimensional manifold. The dilaton depends only on the co-ordinates of $M_{9-n}$, and all $p$-form fields have components only along the $M_{9-n}$ directions. The heterotic warped product $\t{dS}_2$ solutions are the direct product $\t{AdS}_3 \times M_7$ solutions which have been classified in \cite{Beck:2015gqa}. Compared to these types of solutions, the conditions on supersymmetric warped product $\t{dS}_4$ 
solutions in $D=11$ supergravity are rather weaker.

Motivated by these results, in this paper we classify the warped product $\t{dS}_4$ 
solutions in $D=11$ supergravity. We find, on integrating the Killing spinor equations along the $\t{dS}_4$ directions, that all of the necessary and sufficient conditions for supersymmetry are encoded in a single gravitino-type equation, which
is satisfied by a spinor $\psi_+$ whose components depend only on the co-ordinates
of the internal space.  We analyse the solutions of this equation using spinorial geometry techniques. This technique was introduced in \cite{b9} and consists of writing the Killing spinors in terms of multi-differential forms  and, utilizing the gauge-covariance of the KSE, gauge transformations are then used to write the spinors in one of several simple
canonical forms. The main outcome of this approach is a linear system which imposes conditions on the spin connection and the fluxes of the theory. This in turn can be
used to obtain conditions on the geometry which are necessary and sufficient for supersymmetry. These techniques have been applied to classify a wide variety of supergravity solutions \cite{b2}.

In the case of 
warped product $\t{dS}_4$ solutions, we state explicitly the $Spin(7)$ gauge transformations which are used to write the spinor $\psi_+$ in canonical forms with stabilizer subgroups $SU(3)$ and $G_2$. We then solve the linear system obtained from
the Killing spinor equations. In particular, we show that the linear system implies that there are no Killing spinors for which the stabilizer of $\psi_+$ is $G_2$. For the case of $SU(3)$ stabilizer subgroup, the Killing spinor equations determine all components of the 4-form flux in terms of the geometry of the internal manifold, and we present the geometric conditions and the components of the flux, written in a $SU(3)$ covariant fashion. On considering these conditions, we note that the warped product $\t{dS}_4$ geometries are manifestly less restricted in terms of the geometric structure and the 4-form flux in comparison to the warped product $\t{dS}_n$ solutions for $5 \leq n \leq 10$. Our analysis does not utilize the global techniques developed for the investigation of supersymmetric black holes \cite{Gutowski_2013}; we consider only local properties of the Killing spinor equations. This avoids the no-go theorems which exclude warped product $\t{dS}_n$ solutions when the warp product and 4-form flux are smooth, and the internal manifold is smooth and compact without boundary.

The plan of this paper is as follows. In Section 2 we summarize the bosonic field equations, Bianchi identities, and Killing spinor equations of $D=11$ supergravity, we also describe the ansatz for the warped product $\t{dS}_4$ solutions, and present the reduction of the bosonic conditions to the internal manifold. In Section 3, we derive several integrability conditions from the Killing spinor equations, and we demonstrate how some of these integrability conditions can be derived from others.
In Section 4, we explicitly integrate up the Killing spinor equations along the $\t{dS}_4$ directions, and show how the Killing spinor equations reduce to a single gravitino-type equation for a spinor $\psi_+$ which depends only on the internal manifold co-ordinates. We also prove that the supersymmetric $\t{dS}_4$ warped product solutions preserve $N=8n$ supersymmetries for $n=1,2,3,4$. In Section 5 we utilize spinorial geometry techniques, and prove that the spinor $\psi_+$ can be written in a particularly simple canonical form on applying appropriate $Spin(7)$ gauge transformations. Furthermore, we prove that such a spinor has stabilizer subgroup which is either $SU(3)$ or $G_2$; in the $SU(3)$ case we also consider several possible special sub-cases. In Section 6, we present the $SU(3)$ covariant conditions on the flux and geometry, obtained from the gravitino-type equation in the case of $SU(3)$ stabilizer. In Section 7 we also prove that there are no supersymmetric warped product $\t{dS}_4$ solutions for which the stabilizer subgroup of $\psi_+$ is $G_2$. We present our conclusions in Section 8. In Appendix A, we list some conventions. In Appendix B we present some properties of the explicit representation of the Clifford algebras in terms of differential forms, as utilized in the spinorial geometry method. In Appendix C we list some equations which are
used in the process of integrating up the Killing spinor equation along the $\t{dS}_4$ directions in Section 4. In Appendices D and E we list the linear system of equations obtained from the gravitino-type equation for the cases of $\psi_+$ with $SU(3)$ and $G_2$ stabilizer respectively.

\section{Bosonic field equations and KSE}

In this section, we summarize the bosonic field equations and Killing spinor equations (KSE) of $D=11$ supergravity, and describe the ansatz for  the warped product $\t{dS}_4$ geometries. The bosonic fields of $D=11$ supergravity consist of a metric $g$, and a 3-form gauge potential $A$ with 4-form field strength $F=\d A$.
The action for the bosonic fields is given by 
\bea
	S = \frac{1}{2\kappa^2}\int \d^{11}x\sqrt{-g}R-\frac{1}{2}F\wedge *F + \frac{1}{6}F\wedge F\wedge A \ ,
\eea
where $\kappa^2$ is proportional to the gravitational coupling constant.
The equations of motion are thus given by 
\bea
\label{eom}
	& \,R_{AB} -\frac{1}{2}Rg_{AB}-\frac{1}{12}F_{AB_1B_2B_3}F_{B}\,^{B_1B_2B_3}+\frac{1}{96}g_{AB}F^2=0 \nonumber \\
	& \d * F - \frac{1}{2}F\wedge F=0 \ .
\eea
By using 
\bea
\label{rel12}
	R = \frac{1}{144}F^2\  , 
\eea
the first equation in~\eqref{eom} becomes
\bea
\label{rel11}
	R_{AB} -\frac{1}{12}\left(F_{AB_1B_2B_3}F_{B}\,^{B_1B_2B_3}-\frac{1}{12}g_{AB}F^2\right)=0 .
\eea
The supercovariant derivative $\mathcal{D}_A$ is defined as 
\bea
\label{rel2}
	\mathcal{D}_M \equiv \nabla_M -\frac{1}{288}\left(\Gamma_M\,^{A_1A_2A_3A_4} - 8\delta_M^{A_1}\Gamma^{A_2A_3A_4}\right)F_{A_1A_2A_3A_3A_4} \ .
\eea
Bosonic solutions to the equations of motion that preserve at least one supersymmetry are those that admit at least one non-vanishing Killing spinor $\epsilon$, which satisfies
\bea
\label{covconst}
	\mathcal{D}_{A}\epsilon = 0 \ .
\eea	
 
In order to analyse supersymmetric warped product $\t{dS}_4$ solutions, we shall split the $D=11$ spacetime in a 4+7 fashion $\d s^2 = \d \t{S}_4 \times_w M_7$, where $\times_w$ denotes a warped product of $\t{dS}_4$ with an internal manifold $M_7$. 
In terms of the $D=11$ frame, capital latin letters such as A,B denote $D=11$ frame indices. These  D=11 frame indices are split in a 4+7 fashion as follows: we use greek letters for $\t{dS}_4$ frame directions, and latin letters from the middle of the alphabet and onwards for $M_7$. Latin letters from the beginning of the alphabet denote $M_7$ spacetime indices. $M_7$ is equipped with local co-ordinates $y^a$, whereas $\t{dS}_4$ is equipped with local co-ordinates $x^{\mu}$.  For further details about the conventions used are set out in Appendix \ref{conventions}.

The warped $\d \t{S}_4$ product metric $g$ is
\bea
\label{rel1}
	\d s^2 =  A^2\d s^2_{\t{dS}_4} + \d s^2_{M_7} =  \eta_{\mu\nu} {\bf{e}}^\mu {\bf{e}}^\nu + \delta_{ij} {\bf{e}}^i {\bf{e}}^j \  , 
\eea
where the vielbein frame is defined as 
\bea
	\begin{cases}
	{\bf{e}}^\mu &\equiv \  \frac{A}{\mathcal{R}}\d x^{\mu} \\
	{\bf{e}}^i &\equiv  \  e^{i}_{b}\d y^b
	\end{cases}\quad
\eea
with
\bea
	\mathcal{R}(x) = \left(1+\frac{1}{4}K x_{\nu}x^{\nu}\right) \ , \quad \quad x_{\nu} \equiv x^{\alpha}\eta_{\alpha\nu} \ .
\eea
The conformal factor $A$ and the vielbein $e^{j}_{a}$  depend only on $y^a$ co-ordinates. The scalar $K$ is constant and greater than zero.

We require that the field strength $F$  must be invariant under the isometries of $\t{dS}_4$, hence it decomposes as follows:
\bea
\label{rel13}
	F = c {\rm dvol} (\t{dS}_4) + X\ ,
\eea
where $c$ is a constant due to the Bianchi identity and  $X$ is a closed 4-form on $M_7$ depending only on $y^a$ co-ordinates. The gauge field equation ({\ref{eom}}) 
is equivalent to
\bea
\label{rel6}
d (A^4 \star_7 X) = c X \ .
\eea

It will be convenient to state the non-vanishing components of the spin-connection, and curvature components. The non-vanishing spin-connection components are 
\bea
	\Omega_{\mu,\nu\rho} &=& \frac{K}{A}x_{[\nu}\eta_{\rho]\mu} \nonumber \\
	\Omega_{\mu,i\nu}&=& -\frac{\nabla_iA}{A}\eta_{\mu\nu} \nonumber \\
	\Omega_{ijk}&=&\Omega_{ijk}(M_7)\ ,
\eea
where on the LHS Greek indices are frame indices on $\t{dS}_4$, and  on the RHS they are co-ordinate indices on $\t{dS}_4$. $\nabla_i$ denotes the Levi-Civita connection on $M_7$.

The non-vanishing Riemann tensor components are 

\bea
	R_{\mu\nu\alpha\beta} &=& (\eta_{\mu\alpha}\eta_{\beta\nu} - \eta_{\nu\alpha}\eta_{\beta\mu})\left(\frac{K}{A^2}-\frac{\nabla_iA\nabla^iA}{A^2}\right) \nonumber \\
	R_{i\alpha j\beta} &=& -\frac{1}{A}\nabla_i\nabla_jA\,\eta_{\alpha\beta} \nonumber \\
	R_{ijkl} &=& R_{ijkl}(M_7)
\eea
where on the LHS Greek indices are frame indices on $\t{dS}_4$, and  on the RHS they are co-ordinate indices on $\t{dS}_4$. The Ricci curvature tensor components are
\bea
\label{rel14}
R_{\mu\nu} &=& \eta_{\mu\nu}\left(3A^{-2}K - A^{-1}\nabla_i\nabla^iA - 3A^{-2}\nabla_iA\nabla^iA\right) \nonumber \\
	R_{\mu i } &=& 0 \nonumber \\
	R_{ij} & = & -4A^{-1}\nabla_i\nabla_jA+R_{ij}(M_7)
\eea
where on the LHS Greek indices are frame indices on $\t{dS}_4$, and  on the RHS they are co-ordinate indices on $\t{dS}_4$.	

The $(\mu\nu)$-component of the Einstein equations of motion~\eqref{rel11}, imply that 
\bea
\label{rel16}
	3KA^{-1} - \nabla_i\nabla^iA -3A^{-1}\nabla_iA\nabla^iA + \frac{1}{3}c^2A^{-7} + \frac{A}{144}X^2 = 0\ .
\eea
 From the $(ij)$-component of the Einstein equation of motion~\eqref{rel11} and  the third equation in \eqref{rel14}, one finds 
\bea
\label{rel30}
	R_{ij}(M_7) = 4A^{-1}\nabla_i\nabla_jA+\frac{1}{12}X_{ia_1a_2a_3}X_j\,^{a_1a_2a_3}+ \frac{1}{6}c^2A^{-8}\delta_{ij} -\frac{1}{144}X^2\delta_{ij}\ .
\eea
On taking the trace of ({\ref{rel30}}), and using ({\ref{rel6}}) and ({\ref{rel16}}), we obtain
\bea
\label{rel17}
	R(M_7) - 8 A^{-1}\nabla_i\nabla^iA -12 A^{-2}\nabla_iA\nabla^iA +12A^{-2}K
  +\frac{1}{6}c^2A^{-8}- \frac{1}{144}X^2 = 0 \ .
\eea

\section{Integrability Conditions from the KSE}

In this section, we begin the analysis of the KSE by computing the integrability conditions associated with ({\ref{covconst}}). These results will be particularly useful when we explicitly integrate up the KSE along the $\t{dS}_4$ directions in the next section.
From \eqref{rel2}, we find
\bea
\label{rel251}
	{\partial \over \partial x^{\mu}}\epsilon = \frac{1}{\mathcal{R}}\left(-\frac{1}{4}Kx^{\alpha}\Gamma_{\alpha\mu}+\frac{1}{2}\nabla_kA\Gamma^k\Gamma_{\mu}+\frac{A}{288}\Gamma_{\mu}\slashed{X}-\frac{c}{6}A^{-3}\Gamma_{\mu}\tilde{\Gamma}^4\right)\epsilon
\eea
and
\bea
\label{rel252}	
	{\partial \over \partial y^{a}}\epsilon = e^j_a\left(\frac{1}{288}\cancel{\Gamma X}_j+ \frac{c}{12}A^{-4}\Gamma_j\tilde{\Gamma}^4 -\frac{1}{36}\slashed{X}_j-\frac{1}{4}\Omega_{j,lm}\Gamma^{lm}\right)\epsilon\  , 
\eea
where 
\bea
	\tilde{\Gamma}^4 \equiv \Gamma^0\Gamma^1\Gamma^2\Gamma^3 \ . 
\eea
We remark that ({\ref{rel252}}) is equivalent to
\bea
\label{rel33}
		\nabla_i\epsilon =\left(\frac{1}{288}\cancel{\Gamma X}_i+ \frac{c}{12}A^{-4}\Gamma_i\tilde{\Gamma}^4 -\frac{1}{36}\slashed{X}_i\right)\epsilon \ ,
\eea	
where $\nabla_i$ denotes the Levi-Civita connection on $M_7$.

We use these expressions to derive several integrability conditions. First,
from the integrability condition on $\t{dS}_4$ spacetime
\bea
	\left({\partial \over \partial x^{\mu}}{\partial \over \partial x^{\nu}} - {\partial \over \partial x^{\nu}}{\partial \over \partial x^{\mu}}\right) \epsilon= 0\  , 
\eea
we get
\bea
\label{rel5}
	\left(|\nabla A|^2 - K - \frac{c^2}{9}A^{-6} -\frac{A^2}{(144)^2}\slashed{X}^2 +\frac{2}{3}cA^{-3}\nabla_iA\Gamma^i\tilde{\Gamma}^4 -\frac{1}{18}A\nabla_{i}A\slashed{X}^i\right)\epsilon = 0 \ .
\eea
On the other hand, from the  integrability condition with one direction on $\t{dS}_4$ and the other on $M_7$, i.e.
\bea
	\left({\partial \over  \partial x^{\mu}}{\partial \over \partial y^{a}} - {\partial \over  \partial y^{a}}{\partial \over \partial x^{\mu}}\right) \epsilon= 0
\eea
we get
\bea
\label{rel3}
 \Big(\,&-&\frac{1}{2}\nabla_i\nabla_kA\Gamma^{k} +\frac{A}{288}\nabla_i\slashed{X} +\frac{5}{6}\frac{A}{288}\left(\Gamma_{[il_1l_2}\,^{j_3j_4}\,\delta^{j_2}_{l_3}\,\delta^{j_1}_{l_4]}X_{j_1j_2j_3j_4}\,X^{l_1l_2l_3l_4}\right) \nonumber \\
 &+&\frac{5}{6}\frac{A}{288}\left(\Gamma_{[i}X_{l_1l_2l_3l_4]}\,\,X^{l_1l_2l_3l_4}\right) + \frac{c}{864}A^{-3}(10\slashed{X}_i - \Gamma_i\slashed{X})\,\tilde{\Gamma}^4 \nonumber \\
 &+&\frac{A}{144}\Gamma_{l_1}\,^{j_3j_4}\,X_{i}\,^{l_1j_1j_2}\,X_{j_1j_2j_3j_4}+\frac{c}{2}A^{-4}\nabla_iA\,\tilde{\Gamma}^4 +\frac{1}{72}\nabla_kA\Gamma_{l_1l_2l_3i}\,X^{l_1l_2l_3k} \nonumber \\
 &-&\frac{c}{12}A^{-4}\nabla_kA\Gamma^{k}\,_i\,\tilde{\Gamma}^4+\frac{1}{12}\nabla_kA\Gamma^{mn}\,X_{imn}\,^k\Big)\epsilon = 0 \ .
\eea

The integrability conditions ({\ref{rel5}}) and ({\ref{rel3}}) are, however, not independent; (\ref{rel5}) is implied by ({\ref{rel3}}).
To see this, contract \eqref{rel3} with $\Gamma^i$, and using equation of motion~\eqref{rel16} and the Bianchi Identity $\d F = 0$, we are able to derive the integrability condition~\eqref{rel5}. 

So far, we have analyzed the integrability conditions involving the $\t{dS}_4$ part of the covariant derivative~\eqref{rel251}. The integrability condition on $M_7$ given by 
\bea
\label{rel38}
	\left[\nabla_i,\nabla_j\right]\epsilon = \frac{1}{4}R_{ijmn}\Gamma^{mn}\epsilon\  , 
\eea
is 
\bea
\label{intinteg}
  \frac{1}{4}R_{ijmn}\Gamma^{mn}\epsilon\quad &=& \Bigg[ \frac{1}{288}\left(\nabla_i (\cancel{\Gamma X}_j) - \nabla_j (\cancel{\Gamma X}_i)\right) -\frac{c}{3}A^{-5}\left(\nabla_iA\Gamma_j - \nabla_jA\Gamma_i\right)\tilde{\Gamma}^4 \nonumber \\
&-& \frac{1}{36}\left(\nabla_i\slashed{X}_j - \nabla_j\slashed{X}_i\right) +\frac{1}{288^2}\left(\cancel{\Gamma X}_j\cancel{\Gamma X}_i -\cancel{\Gamma X}_i\cancel{\Gamma X}_j\right)
\nonumber \\
&+&\frac{1}{36^2}\left(\slashed{X}_j\slashed{X}_i-\slashed{X}_i\slashed{X}_j\right) 
+ \frac{1}{288}\frac{c}{12}A^{-4}\left(\cancel{\Gamma X}_j\Gamma_i 
- \Gamma_i \cancel{\Gamma X}_j\right)\tilde{\Gamma}^4 
\nonumber \\
&+& \frac{1}{288}\frac{c}{12} A^{-4}\left(\Gamma_j\cancel{\Gamma X}_i-\cancel{\Gamma X}_i\Gamma_j\right) \tilde{\Gamma}^4 
+ \frac{1}{288}\frac{1}{36}\left(\cancel{\Gamma X}_i\slashed{X}_j - \slashed{X}_j \cancel{\Gamma X}_i\right)
\nonumber \\
 &+& 
\frac{1}{288}\frac{1}{36}\left(\slashed{X}_i\cancel{\Gamma X}_j - \cancel{\Gamma X}_j\slashed{X}_i\right) 
+ \frac{c^2}{72}A^{-8}\Gamma_{ij} 
\nonumber \\
&+& \frac{c}{432}A^{-4}\left(\slashed{X}_i\Gamma_j - \Gamma_j\slashed{X}_i\right)\tilde{\Gamma}^4  
+\frac{c}{432}A^{-4} \left(\Gamma_i\slashed{X}_j - \slashed{X}_j\Gamma_i\right)\tilde{\Gamma}^4 \Bigg]\epsilon \ .
\nonumber \\
\eea

In fact, ({\ref{rel3}}) is implied by ({\ref{intinteg}}). To see this, contract
({\ref{intinteg}}) with $\Gamma^j$ and use the Einstein equation \eqref{rel30}, the
Bianchi identity, $R_{l[ijk]} = 0$, and the condition $dX=0$, as well as the gauge field equations ({\ref{rel6}}). 
In particular:
\begin{itemize}
\item The condition $\d X = 0 $ is used to derive:
\bea
4\Gamma^{la_1a_2a_3}\nabla_lX_{ka_1a_2a_3} = \nabla_k\slashed{X} \ .
\eea 
\item The gauge field equation ~\eqref{rel6} implies that
 from 
\bea
4A(\Gamma_i\nabla_j\slashed{X}^j - \nabla_j\slashed{X}^j\Gamma_i ) + cA^{-3}(\Gamma_i\slashed{X} - \slashed{X}\Gamma_i)\,\tilde{\Gamma}^4
\nonumber \\
+16\nabla_jA(\Gamma_i\slashed{X}^j-\slashed{X}^j\Gamma_i)=0 \ ,
\nonumber \\
\eea
and from this condition, it follows that
	\bea
	A\Gamma_{ia_1a_2a_3}\nabla_kX^{ka_1a_2a_3} = -cA^{-3}\slashed{X}_i\tilde{\Gamma}^4 - 4\nabla_kA\Gamma_{ia_1a_2a_3}X^{ka_1a_2a_3} \ .
	\eea
	\item The gauge field equation  ~\eqref{rel6} also implies that 
\bea
4A(\Gamma_i\nabla_j\slashed{X}^j + \nabla_j\slashed{X}^j\Gamma_i ) + cA^{-3}(\Gamma_i\slashed{X} + \slashed{X}\Gamma_i)\,\tilde{\Gamma}^4
\nonumber \\
+16\nabla_jA(\Gamma_i\slashed{X}^j+\slashed{X}^j\Gamma_i)=0 \ ,
\nonumber \\
\eea
and from this condition, it follows that
\bea
	A\Gamma^{ab}\nabla^kX_{kiab} = -\frac{1}{12}cA^{-3}\Gamma_i\slashed{X}\tilde{\Gamma}^4 +\frac{1}{3}cA^{-3}\slashed{X}_i\tilde{\Gamma}^4 - 4 \nabla^kA\Gamma^{ab}X_{kiab} \ .
\eea
\end{itemize}

Hence, it follows that the integrability conditions ({\ref{rel5}}) and ({\ref{rel3}})
are both implied by ({\ref{intinteg}}), which is derived from the integrability condition of ({\ref{rel33}}).

\section{Integration of KSE}

In this section, we will explicitly integrate the KSE along the $\t{dS}_4$ directions. In this analysis, we shall show that the KSE reduce to a single gravitino-type KSE acting on a spinor $\psi$ which is independent of the $\t{dS}_4$ co-ordinates.
To begin, we shall define a spinor $\Phi$, as follows:

\bea
\label{rel37}
	\Phi \equiv \frac{A}{288}\slashed{X}\epsilon -\frac{1}{2}\nabla_kA\Gamma^k\epsilon + acA^{-3}\tilde{\Gamma}^4\epsilon\  , 
\eea
where $a$ is a constant to be fixed. We have chosen the relative coefficients between $\slashed{X}$  and $\slashed{\d A}$ in \eqref{rel37} motivated by the first two terms in \eqref{rel3}. We shall show that one can choose the constant $a$, as well as other constants $k_1$, $k_2$, $q_1$, $q_2$, $q_3$, $q_4$, $q_5$ such that
\bea
\label{rel32}
	&&\nabla_i\Phi +k_1 [\t{Eq.}~\eqref{rel3}] + k_2A^{-1}\Gamma^i[\t{Eq.}~\eqref{rel5}] \nonumber \\
	&& + q_1 \cancel{\Gamma X}_i\Phi + q_2\slashed{X}_i\Phi + q_3cA^{-4}\Gamma_i\tilde{\Gamma}^4\Phi + q_4A^{-1}\nabla_kA\Gamma_i\Gamma^k\Phi + q_5 A^{-1}\nabla_iA\Phi = 0\  .
\nonumber \\
\eea
Details of this calculation are presented in Appendix~\ref{appendix_constant}. One finds that 
\bea
	k_1 = -1 \quad a  = -\frac{1}{6} \quad q_1  = \frac{1}{288} \quad q_2  = -\frac{1}{36}  \quad q_3  = -\frac{1}{12}\quad k_2  =q_4= q_5 = 0\ . \
\eea
Given this choice of constants, the spinor $\Phi$ is 
\bea
\label{rel21}
	\Phi = \left(\frac{A}{288}\slashed{X}-\frac{1}{2}\nabla_kA\Gamma^k-\frac{c}{6}A^{-3}\tilde{\Gamma}^4\right)\epsilon\  , 
\eea
which  satisfies the following equations

\bea
\label{rel201}
	{\partial \over \partial x^{\mu}}\Phi =\, \frac{1}{\mathcal{R}}\left[-\frac{1}{4}Kx^{\alpha}\Gamma_{\alpha\mu}+\Gamma_{\mu}\left(\frac{1}{2}\nabla_kA\Gamma^k+\frac{A}{288}\slashed{X}+\frac{c}{6}A^{-3}\tilde{\Gamma}^4\right)\right]\Phi
\eea
\bea
\label{rel202}
	{\partial \over \partial y^{a}}\Phi =\,e^j_a\left(-\frac{1}{288}\cancel{\Gamma X}_j+ \frac{c}{12}A^{-4}\Gamma_j\tilde{\Gamma}^4 +\frac{1}{36}\slashed{X}_j-\frac{1}{4}\Omega_{j,lm}\Gamma^{lm}\right)\Phi\quad .
\eea
These equations are similar, but not identical, to the original Killing spinor equations
for $\epsilon$~\eqref{rel251}-\eqref{rel252}. The differences are in terms of certain signs appearing in \eqref{rel201}-\eqref{rel202}, which
are flipped with respect to \eqref{rel251}-\eqref{rel252} - in~\eqref{rel201}  the second and the fourth term with respect to~\eqref{rel251} and in~\eqref{rel202}  the 
first and the third term with respect to ~\eqref{rel252}.

Equations~\eqref{rel201} and \eqref{rel202}, will be particularly useful in the process of integrating up the KSE along the $\t{dS}_4$ directions.
By using \eqref{rel5}, \eqref{rel201} becomes 
\bea
\label{rel23}
	{\partial \over \partial x^{\mu}}\Phi = -\frac{K}{4\mathcal{R}}x^{\alpha}\Gamma_{\alpha\mu}\Phi -\frac{K}{4\mathcal{R}}\Gamma_{\mu}\epsilon\ . \
\eea
By using the definition of $\Phi$~\eqref{rel21}, one can rewrite ${\partial \over \partial x^{\mu}}\epsilon$ as 
\bea
\label{rel22}
	{\partial \over \partial x^{\mu}}\epsilon  = \frac{1}{\mathcal{R}}\left(-\frac{1}{4}Kx^{\alpha}\Gamma_{\alpha\mu}\epsilon + \Gamma_{\mu}\Phi \right)\ . \
\eea
Applying a second derivative ${\partial \over \partial x^{\nu}}$ to \eqref{rel22}, using \eqref{rel23} and finally exploiting \eqref{rel22} to cancel $\mathcal{R}^{-1}\Gamma_{\mu}\Phi$ terms, one gets a second order differential equation for $\epsilon$, namely 
\bea
\label{rel19}
	{\partial \over \partial x^{\mu}}{\partial \over \partial x^{\nu}}\epsilon +\frac{K}{4\mathcal{R}}(x_{\mu}{\partial \over \partial x^{\nu}}\epsilon + x_{\nu}{\partial \over \partial x^{\mu}}\epsilon) -\frac{K^2}{16\mathcal{R}^2}x_{\mu}x_{\nu}\epsilon +\frac{K}{4\mathcal{R}}\eta_{\mu\nu}\epsilon = 0\ . \
\eea
On defining $\eta$ by
\bea
\label{rel24}
	\epsilon = \mathcal{R}^{-\frac{1}{2}} \eta \ ,
\eea
it is straightforward to see that ({\ref{rel19}}) is equivalent to
\bea
	{\partial \over \partial x^{\mu}}{\partial \over \partial x^{\nu}}\eta = 0\quad\Rightarrow\quad \eta = \psi + x^{\lambda}\tau_{\lambda}\ ,
\eea
and hence this equation can be integrated to find
\bea
\eta = \psi + x^{\lambda}\tau_{\lambda} \ ,
\eea
where $\psi$, $\tau_{\lambda}$ with $\lambda = 0,1,2,3$ are Majorana spinors which do not depend on the $x_{\mu}$ co-ordinates.

Given this  expression for $\epsilon$, i.e.
\bea
\label{finaleps}
	\epsilon =  \mathcal{R}^{-\frac{1}{2}}(\psi + x^{\lambda}\tau_{\lambda})\  , 
\eea
we substitute it into the KSEs~\eqref{rel251} and \eqref{rel252}. As the spinors 
$\psi$, $\tau_\lambda$ are independent of the $\t{dS}_4$ co-ordinates, on expanding
\eqref{rel251} and \eqref{rel252} order-by-order in $x_\alpha$, we find various conditions.

In particular, from the KSE along the $\t{dS}_4$ directions \eqref{rel251}, the vanishing of $x-$independent terms imply that the Majorana spinors $\tau_{\mu}$ are given in terms of $\psi$, as follows:
\bea
\label{rel26}
	\tau_{\mu} = \Gamma_{\mu}\left(\frac{A}{288}\slashed{X}-\frac{1}{2}\nabla_kA\Gamma^k - \frac{c}{6}A^{-3}\tilde{\Gamma}^4 \right)\psi\ . \
\eea

The vanishing of the terms that are linear in $x_{\mu}$ in  \eqref{rel251} imply
\bea
\label{rel28}
	\left(|\nabla A|^2 - K - \frac{c^2}{9}A^{-6} -\frac{A^2}{(144)^2}\slashed{X}^2 +\frac{2}{3}cA^{-3}\nabla_iA\Gamma^i\tilde{\Gamma}^4 -\frac{1}{18}A\nabla_{i}A\slashed{X}^i\right)\psi = 0 \ ,
\eea
and we remark that this condition is equivalent to the integrability condition
\eqref{rel5}, but with $\epsilon$ replaced with $\psi$. The terms in \eqref{rel251} 
which are quadratic in $x_{\mu}$ vanish identically; this then exhausts the content of
\eqref{rel251}. 

Next we consider the  KSE along the seven-dimensional internal directions,
\eqref{rel252}. Again, we substitute in ({\ref{finaleps}}) and expand order-by-order in $\t{dS}_4$ co-ordinates. The vanishing of $x-$independent terms gives
\bea
\label{rel33b}
		\nabla_i\psi =\left(\frac{1}{288}\cancel{\Gamma X}_i+ \frac{c}{12}A^{-4}\Gamma_i\tilde{\Gamma}^4 -\frac{1}{36}\slashed{X}_i\right)\psi\ . \
\eea	
The above equation~\eqref{rel33b} implies that $\psi$ satisfies a gravitino KSE along the internal directions, which is identical to the condition 
\eqref{rel33b} but with $\epsilon$ replaced with $\psi$.

From the terms in \eqref{rel252} which are linear in $x_\mu$ we obtain

\bea
\label{rel29}
	&&\Big[ \frac{A}{288}\nabla_i \slashed{X} - \frac{1}{2}\nabla_i\nabla_kA\Gamma^k +\frac{A}{1728}\Gamma_iX^2 +\frac{A}{864}\Gamma_{j_1j_2j_3}\,^{l_1l_2}X_{ij_4l_1l_2}X^{j_1j_2j_3j_4} 
\nonumber \\
	&-& \frac{A}{432}\Gamma_{j_1}X_{ij_2j_3j_4}X^{j_1j_2j_3j_4} -\frac{A}{576}\Gamma_{ij_1j_2}\,^{l_1l_2}X_{j_3j_4l_1l_2}X^{j_1j_2j_3j_4} \nonumber \\
	&-& \frac{1}{864}cA^{-3}\Gamma_i\slashed{X}\tilde{\Gamma}^4 + \frac{5}{432}cA^{-3}\slashed{X}_i\tilde{\Gamma}^4+ \frac{A}{144}\Gamma^m\,_{ab}X_{impq}X^{pqab} \nonumber \\
	&+& \frac{1}{72}\nabla_kA\Gamma_{ij_1j_2j_3}X^{kj_1j_2j_3}+ \frac{1}{12}\nabla_kA\Gamma^{ab}X_{i}\,^{k}\,_{ab} \nonumber \\
	 &-&\frac{c}{12}A^{-4}\nabla_kA\Gamma^k\Gamma_i\tilde{\Gamma}^4 + \frac{7}{12}cA^{-4}\nabla_iA \tilde{\Gamma}^4        \Big]\psi = 0
\eea
which is identical to the integrability condition \eqref{rel3},
with $\epsilon$ replaced by $\psi$. This then exhausts the content of
\eqref{rel252}. 

Hence, we have shown that the spinor $\epsilon$ is given by
\bea
\label{d11ksp}
	\epsilon =  \mathcal{R}^{-\frac{1}{2}}(\psi + x^{\lambda}\tau_{\lambda})\ ,
\eea
where
\bea
	\tau_{\lambda} = \Gamma_{\lambda}\left(\frac{A}{288}\slashed{X}-\frac{1}{2}\nabla_kA\Gamma^k - \frac{c}{6}A^{-3}\tilde{\Gamma}^4 \right)\psi \ .
\eea
The Majorana spinor $\psi$ is independent of the $\t{dS}_4$ co-ordinates, and satisfies 
({\ref{rel33b}}). Furthermore, $\psi$ must also satisfy the algebraic conditions ({\ref{rel29}}) and
({\ref{rel28}}). However, as we have shown in the previous section, the integrability conditions of ({\ref{rel33b}}), together with the bosonic field equations and Bianchi identities, imply that ({\ref{rel29}}) holds. Furthermore, we have shown that ({\ref{rel29}}) also implies ({\ref{rel28}}). Hence, the necessary and sufficient conditions for supersymmetry are encoded in ({\ref{rel33b}}).

\subsection{Counting the supersymmetries}

Having determined that the necessary and sufficient conditions for supersymmetry
are given by ({\ref{rel33b}}), we shall now count the number of solutions to this equation.  In particular, if $\psi$ satisfies ({\ref{rel33b}}), then so does $\Gamma_{\mu\nu}\psi$.
We choose a null basis for the Majorana representation of $\t{Spin(10,1)}$ and
take the $\t{dS}_4$ frame directions to correspond with the ${+,-,1,\bar{1}}$ directions, see Appendix~\ref{hSpinorsfromforms}. 
The frame directions associated with the internal manifold $M_7$ correspond to the  $2,3,4,\bar{2},\bar{3},\bar{4},\#$ directions.

With these conventions for the de Sitter and internal manifold frames, we define lightcone projection operators as 
\bea
	P_{\pm} \equiv \frac{1}{2}\left(\mathbb{I} \pm \Gamma_{+-}\right)\ .
\eea
As the projection operator $P_{\pm}$ commutes with the supercovariant derivative~\eqref{rel33b}, we then decompose the spinor $\psi$ using the lightcone projectors and we define $\psi_\pm$ to be 
\bea
	\psi_\pm \equiv P_\pm \psi \quad\Rightarrow\quad \Gamma_\pm \psi_\pm = 0 \ . 
\eea

Without loss of generality, utilizing these projection operators, any supersymmetric
solution must admit a positive chirality solution $\psi_+$ to ({\ref{rel33b}}).
Given such a $\psi_+$ spinor, we can then define
\bea
\tilde{\psi}_+  &\equiv& i \Gamma_{1\bar{1}}\psi_+ \nonumber\\
\psi_-  &\equiv& \Gamma_-(\Gamma_1 + \Gamma_{\bar{1}})\psi_+ \nonumber\\
\tilde{\psi}_- &\equiv& i\Gamma_-(\Gamma_1 - \Gamma_{\bar{1}})\psi_+ \ .
\eea

$\tilde{\psi}_+$ is an additional positive chirality solution to ({\ref{rel33b}}), and
$\{ \psi_- , \tilde{\psi}_- \} $ are two  negative chirality
solutions to ({\ref{rel33b}}). $\{ \psi_+, \tilde{\psi}_+ , \psi_- , \tilde{\psi}_- \}$
are linearly independent, as by construction they are mutually orthogonal with respect to the Dirac inner product $\langle~\cdot~,~\cdot~\rangle$.

It would therefore appear, a priori, that the number of supersymmetries is $4n$. However, there are, in fact further additional spinors. To see this, note that
({\ref{rel202}}) implies that 
\bea
	\check{\psi}_+ \equiv (\Gamma_1+\Gamma_{\bar{1}})\left(\frac{A}{288}\slashed{X}-\frac{1}{2}\nabla_kA\Gamma^k-\frac{c}{6}A^{-3}\tilde{\Gamma}^4\right)\psi_+\  , 
\eea
is also a positive chirality solution of ({\ref{rel33b}}). Furthermore, it can be shown that $\{ \psi_+, \tilde{\psi}_+ , \check{\psi} \}$ are linearly independent. To see this, suppose that
\bea
\label{linind1}
\check{\psi}_+ = c_1 \psi_+ + i c_2 \Gamma_{1 \bar{1}} \psi_+ \ ,
\eea
for real constants $c_1$, $c_2$. Acting on both sides of this condition with the operator
$(\Gamma_1 + \Gamma_{\bar{1}}) \big(\frac{A}{288}\slashed{X}-\frac{1}{2}\nabla_kA\Gamma^k-\frac{c}{6}A^{-3}\tilde{\Gamma}^4 \big)$, and utilizing
the integrability condition ({\ref{rel28}}) to simplify the LHS, we find
\bea
-{K \over 2} \psi = (c_1^2+ c_2^2) \psi \ ,
\eea
where we have also used ({\ref{linind1}}) to simplify the RHS. It is clear that this
admits no solution, as $K>0$. Hence, we find that we can construct four linearly independent positive chirality spinors which solve ({\ref{rel33b}}), corresponding to $\{ \psi_+, \tilde{\psi}_+ , \check{\psi}_+, {\tilde{\check{\psi}}}_+ \}$,
where ${\tilde{\check{\psi}}}_+ \equiv i \Gamma_{1 \bar{1}} \check{\psi}$. There are
also four corresponding negative chirality spinors given by
$\{\psi_-, {\tilde{\psi}}_-, {\check{\psi}}_-, {\tilde{\check{\psi}}}_- \}$, where
\bea
{\check{\psi}}_- = \Gamma_- (\Gamma_1+\Gamma_{\bar{1}}) {\check{\psi}}_+ \ ,
\qquad {\tilde{\check{\psi}}}_- = i \Gamma_{1 \bar{1}} {\check{\psi}}_- \ .
\eea

Hence we have constructed 8 linearly independent solutions to ({\ref{rel33b}}),

\bea
\{ \psi_+, \tilde{\psi}_+ , \check{\psi}_+, {\tilde{\check{\psi}}}_+ , \psi_-, {\tilde{\psi}}_-, {\check{\psi}}_-, {\tilde{\check{\psi}}}_-  \}
\eea
and it follows that the number of supersymmetries for warped product $\t{dS}_4$ solutions is $8n$, $n=1,2,3,4$. 

We remark that the existence of the additional spinors ${\check{\psi}}_\pm$, ${\tilde{\check{\psi}}}_\pm$ is somewhat analogous to results found in the analysis of near-horizon geometries of supersymmetric extremal black holes~\cite{Gutowski_2013} and also for warped product $\t{AdS}$ solutions \cite{b1}.
In these cases, given a Killing spinor, one also finds that additional Killing spinors can be generated by the action of certain algebraic operators constructed out of the fluxes of the theory.

\section{Spinorial Geometry: Canonical forms for $\psi$}
\label{Canonical}

In this section we shall use $Spin(7)$ gauge transformations to  bring the spinor $\psi_+$ to one of several simple canonical forms. We will describe the gauge transformations used to do this explicitly.

 The most general form of a positive chirality Majorana spinor $\psi_+ \in \Delta_{32}$  can be expressed by using \eqref{spinorsappendix1}, i.e.
\bea
	\psi_+ &=& w 1 + \bar{w}\t{e}_{1234}   + \lambda^{1}\t{e}_{1} + \bar{\lambda}^1\t{e}_{234}+ \lambda^j\t{e}_{j}-\frac{1}{3!}(*\bar{\lambda})^{l_1l_2l_3}\t{e}_{l_1l_2l_3}  
\nonumber\\
&+& \Omega^{q}\t{e}_{1q} -\frac{1}{2!}\bar{\Omega}^{q}\varepsilon_{q}\,^{mn}\t{e}_{mn}\ ,
\nonumber \\
\eea
with $l, q, m, m = 2,3,4$.
As the action of $\t{SU}(N)$ on $\mathbb{C}^N-\{0\}$ is transitive and the generating orbits are $(2N-1)$-spheres, one can apply a SU(3) gauge transformation in the 2,3,4 directions to set, without loss of generality, $\Omega^3 = \Omega^4 = 0 $~\footnote{Generally, the complex value $\Omega^2$ can be set to be real with the same SU(3) transformation used to set $\Omega^3 = \Omega^4 = 0 $. It does not happen in this specific case due to the fact that $\Omega^2$ will be promoted to be  complex value in the next gauge transformation.}  i.e.
\bea
	\psi_+ = w 1 + \bar{w}\t{e}_{1234} +  \lambda^{1}\t{e}_{1} + \bar{\lambda}^1\t{e}_{234}+\lambda^j\t{e}_{j}-\frac{1}{3!}(*\bar{\lambda})^{l_1l_2l_3}\t{e}_{l_1l_2l_3}  + \Omega\t{e}_{12} - \bar{\Omega}\t{e}_{34} \ .
\eea
To proceed further, we define $T^{1},T^2,T^3$ as
\bea
	T_1 \equiv \frac{1}{2}(\Gamma_{34} + \Gamma_{\bar{3}\bar{4}})\quad T_2 \equiv \frac{i}{2}(\Gamma_{34} - \Gamma_{\bar{3}\bar{4}})\quad T_3 \equiv T_1T_2
= {i \over 2} (\Gamma_{3 \bar{3}}+ \Gamma_{4 \bar{4}}) \ .
\eea
It is straightforward to verify that $T^i$ with $i=1,2,3$, which satisfy the algebra of the imaginary unit quaternions, preserve the span of the following basis elements
\bea
	v_1 \equiv&\, (1+\t{e}_{1234}) \qquad v_2  \equiv i(1- \t{e}_{1234}) \nonumber \\
	v_3 \equiv&\, (\t{e}_{12}-\t{e}_{34}) \qquad v_4 \equiv         i(\t{e}_{12}+\t{e}_{34}) 
\eea
and we remark that the Spin(7) gauge transformation generated by the $T_i$ is of the form
$p^4 {\rm id} + p^i T_i$ where $(p^1, p^2, p^3, p^4) \in S^3$.

Then one can carry out a SO(2) gauge transformation generated by $T_3$ to set $w \in \mathbb{R}$. So far, the spinor $\psi_+$ can be written as
\bea
	\psi_+ = w (1 +\t{e}_{1234}) + \lambda^{1}\t{e}_{1} + \bar{\lambda}^1\t{e}_{234} +\lambda^j\t{e}_{j}-\frac{1}{3!}(*\bar{\lambda})^{l_1l_2l_3}\t{e}_{l_1l_2l_3}  + \Omega\t{e}_{12} - \bar{\Omega}\t{e}_{34}\ .
\eea
A SU(3) gauge transformation generated by $i(\Gamma_{2\bar{2}} -{1 \over 2}\Gamma_{3\bar{3}} -{1 \over 2}\Gamma_{4\bar{4}} )$, which leaves
$\{1, e_{1234} \}$ invariant, is then used to set $\Omega \in \mathbb{R}$, so
\bea
	\psi_+ = w (1 +\t{e}_{1234}) + \lambda^{1}\t{e}_{1} + \bar{\lambda}^1\t{e}_{234} + \lambda^j\t{e}_{j}-\frac{1}{3!}(*\bar{\lambda})^{l_1l_2l_3}\t{e}_{l_1l_2l_3}  + \Omega(\t{e}_{12} -\t{e}_{34})\ .
\eea
 	
We next exploit a SO(2) transformation generated by $T_1$, acting on $v_1$ and $v_3$ to put $\Omega = 0$.	
Then, we make a further SU(3) gauge transformation along the $2,3,4$ directions to set $\lambda^3=\lambda^4 = 0$ with $\lambda^2 \in \mathbb{R}$, i.e. 
\bea
\label{orbsimp1a}
	\psi_+ = w (1 +\t{e}_{1234}) + \lambda^{1}\t{e}_{1} + \bar{\lambda}^1\t{e}_{234} + \lambda^2(\t{e}_{2} - \t{e}_{134})\ . \
\eea
In order to simplify further the spinor $\psi_+$, 
we shall introduce additional $Spin(7)$ generators $L_1, L_2, L_3$ given by
\bea
L_1 \equiv {1 \over \sqrt{2}} \Gamma_\sharp (\Gamma_2 + \Gamma_{\bar{2}}),
\quad L_2 \equiv {i \over \sqrt{2}} \Gamma_\sharp (\Gamma_2 - \Gamma_{\bar{2}}),
\quad L_3 \equiv L_1 L_2 = i \Gamma_{2 \bar{2}} \ .
\eea
The $L_j$ also satisfy the algebra of the imaginary unit quaternions, and commute with
the $T_i$, and the $Spin(7)$ gauge transformation generated by the $L_j$ is of the form
$q^4 {\rm id} + q^j L_j$ where $(q^1, q^2, q^3, q^4) \in S^3$. We shall then consider a generic gauge transformation generated by the $T_i$ and $L_j$ of the acting on the spinor ({\ref{orbsimp1a}}) of the form
\bea
\label{orbsimp1b}
\big(p^4 {\rm id} + p^i T_i \big) \big(q^4 {\rm id} + q^j L_j \big) \psi_+ \ .
\eea
We set $q^2=q^3=0$ and $q^1 = \sin \sigma, q^4 = \cos \sigma$, such that
\bea
w \lambda^2 \cos 2 \sigma + {1 \over 2} \sin 2 \sigma (\omega^2 - (\lambda^2)^2
-|\lambda^1|^2) =0
\eea
and 
\bea
p^1 &=& \ell {\rm Re} (\lambda^1) \sin \sigma, \quad
p^2 = - \ell {\rm Im} (\lambda^1) \sin \sigma, \nonumber \\
p^3 &=& 0, \quad \quad \quad \quad \quad \quad ~ p^4 = \ell (\omega \cos \sigma - \lambda^2 \sin \sigma)
\eea
where the constant $\ell$ is chosen such that $(p^1, p^2, p^3, p^4) \in S^3$. With this
choice of parameters, the gauge transformation given in ({\ref{orbsimp1b}}) can be used to set $\lambda^2=0$ in ({\ref{orbsimp1a}}), so the simplest canonical form for the spinor $\psi_+$ is given by
\bea
\label{rel39}
	\psi_+ = w (1 +\t{e}_{1234}) + \lambda\t{e}_{1} + \bar{\lambda}\t{e}_{234} \qquad w \in \mathbb{R}, \ \lambda \in \mathbb{C} \ .
\eea

\subsection{Stabilizer Group of $\psi_+$ }
It is useful to consider the stabilizer subgroup of $Spin(7)$ which leaves 
$\psi_+$ invariant. In particular, we must determine the generators $f^{ij}\Gamma_{ij}$,
where $f^{ij} \in \mathbb{R}$ are antisymmetric in $i,j$, and 
satisfy
\bea
\label{rel40}
    f^{ij}\Gamma_{ij}\psi_+ =  0\qquad i,j=\#,\alpha,\bar{\alpha}\  .
\eea
The conditions obtained from \eqref{rel40} are
\bea
\label{stabeq}
    2wf^{\alpha\beta} &=&  \sqrt{2}\bar{\lambda}f^{\#\bar{\rho}}\varepsilon_{\bar{\rho}}\,^{\alpha\beta} \nonumber \\
     2\lambda f^{\alpha\beta} &=&  \sqrt{2}wf^{\#\bar{\rho}}\varepsilon_{\bar{\rho}}\,^{\alpha\beta} \nonumber \\
    f^{\alpha}\,_{\alpha} &=& 0 \ .
\eea
Depending on $w$ and $\lambda$, there are two possible different stabilizer subgroups:

\begin{itemize}
	\item[(a)] if $w^2-|\lambda|^2 \neq 0$ then ({\ref{stabeq}}) implies that $f_{\alpha \beta}=0$ and $f_{\sharp \alpha}=0$, hence the stabilizer is $SU(3)$.
	
	\item[(b)] if $w^2-|\lambda|^2 = 0$, the stabilizer is $G_2$.

\end{itemize} 

In the $SU(3)$ stabilized case it is particularly useful to consider the complex $SU(3)$ invariant spinor bilinear scalar $\langle \psi_+, \Gamma_{\bar{1}} \psi_+ \rangle = 2 \sqrt{2} w \lambda$. There are various different cases, corresponding to whether this scalar vanishes, or it does not vanish:

\begin{itemize}
	\item[(i)] $w \neq 0$, $\lambda = 0$ ,
	
	\item[(ii)] $\lambda \neq 0$, $w = 0$ ,
	
	\item[(iii)] $w \neq  0 \t{ , }  \lambda \neq 0$ .
\end{itemize} 

In fact, it is straightforward to see that the spinors associated with cases $(i)$ and $(ii)$ above are related by a $Pin(7)$ transformation. To see this, consider
the spinor from case $(ii)$,
\bea
	\psi_+ = \lambda \t{e}_1 + \bar{\lambda}\t{e}_{234} \ .
\eea
The $Spin(7)$ gauge transformation generated by $L_3$ produces a $SO(2)$ which
acts transitively on $\{e_1+e_{234}, i(e_1 -e_{234}) \}$, and hence without loss of generality we can set $\psi_+ = \lambda (e_1 + e_{234})$ for $\lambda \in \mathbb{R}$. Next, note that
\bea
\Gamma_{234} (e_1 + e_{234}) = -(1+e_{1234}) \ .
\eea
It therefore follows that the spinor $\psi_+$ in case (ii) is 
$Spin(7)$ gauge-equivalent to a spinor which in turn is Pin-equivalent, with respect to $\Gamma_{234} \in Pin(7)$, to the spinor in case (i). The effect of the
$\Gamma_{234}$ transformation is to flip holomorphic with anti-holomorphic directions  and to reflect along the $\#$ direction, namely
\bea
    \alpha \ \rightarrow \ \bar{\alpha}\qquad , \quad \ \#  \ \rightarrow \ - \# \ .
\eea

It is therefore sufficient to consider spinors $\psi_+$ corresponding to 
the $G_2$ stabilizer case, and the two $SU(3)$ stabilizer cases $(i), (iii)$. Having determined the stabilizers associated with these three canonical types of
spinors, we next proceed to obtain a linear system of equations by substituting these
expressions for $\psi_+$ into \eqref{rel33b}. The linear system consists of relations between the flux and spin-connection, which when covariantized with respect to the appropriate stabilizer group, give rise to conditions on the flux $X$ 
and the geometry of the internal manifold $M_7$. In the following sections, we shall present the covariant solution of the linear system for each of the stabilizer subgroups.

\section{SU(3) Invariant Spinor}

In this section, we solve the KSEs~\eqref{rel33b} when the stabilizer of $\psi_+$ is $SU(3)$, corresponding to 
\bea
	\psi_+ = w (1 +\t{e}_{1234}) + \lambda\t{e}_{1} + \bar{\lambda}\t{e}_{234} \qquad w \in \mathbb{R}, \ \lambda \in \mathbb{C} \ ,
\eea
for $w^2-|\lambda|^2 \neq 0$. We begin by considering the case for which both $w$ and $\lambda$ are non-vanishing. Furthermore, we will write $\lambda = \rho e^{i\theta}$, where $\rho >0 $ and $\theta \in [0,2\pi[$ are two real spacetime functions. The associated linear system   and the components of the flux are presented in Appendix~\ref{linearsystem}.  The linear system is initially expressed non-covariantly in terms of SU(3)-components of the spin-connection and the fluxes, but, it can be rewritten in SU(3)-covariant form by  using the $\t{SU}(3)$ gauge invariant bilinears. 
In Appendix~\ref{covariantexpression} we set out the main relations which are used to write the relations in a manifestly $SU(3)$ covariant fashion, in terms of the following $SU(3)$ invariant bilinears:
\bea
    \xi\equiv \t{e}^{\#} \ , \qquad \omega \equiv -i\delta_{\alpha\bar{\beta}}\,\t{e}^{\alpha}\wedge\t{e}^{\bar{\beta}} \ , \qquad\chi\equiv \frac{1}{3!}\varepsilon_{\alpha\beta\gamma}\t{e}^{\alpha}\wedge\t{e}^{\beta}\wedge\t{e}^{\gamma}\ . \
\eea
The above forms are obtained from the following $SU(3)$-invariant spinor bilinears:
\bea
\langle \psi_+,\Gamma_a\psi_+\rangle \t{e}^a &=& -2(w^2-|\lambda|^2)\xi  
\eea
\bea
{1 \over 2!}\langle \psi_+,\Gamma_{ab}\tilde{\Gamma}^4\psi_+\rangle \t{e}^{a}\wedge\t{e}^b &=& -2(w^2-|\lambda|^2)\omega 
\eea
\bea
{1 \over 3!}\langle \psi_+, (\Gamma_1 + \Gamma_{\bar{1}}) \Gamma_{abc} \psi_+ \rangle \t{e}^a\wedge\t{e}^b \wedge\t{e}^c&=& 2iw(\bar{\lambda}-\lambda)\xi\wedge\omega  
\nonumber \\
&+&4(w^2-\lambda^2)\chi +4(w^2-\bar{\lambda}^2)\bar{\chi}\nonumber \\ 
\eea
where  $a,b,c = \alpha,\bar{\alpha},\#$.

In constructing the solution to the linear system~\eqref{grel1}-\eqref{grel2}, it is convenient to make use of the two Lee forms built from $\chi$, and $\omega$, which are 
\bea
    Z_i \equiv  \nabla^j\chi_{jkl}\bar{\chi}^{kl}\,_i \ , \qquad  W_{i} \equiv \nabla^j\omega_{jk} \omega^{k}\,_{i} \ .
\eea
After some computation, the SU(3)-covariant conditions are as follows (here $i,j,k = \#,\alpha,\bar{\alpha}$ are frame indices on $M_7$):
\bea
    \nabla^i\xi_i = -\frac{6\xi^j}{(w^2-|\lambda|^2)^2} \Big[w(2w^2+3|\lambda|^2)(\d w)_j +(3w^2+2|\lambda|^2)\mathfrak{Re}(\lambda\d\bar{\lambda})_j\Big]
\eea

\bea
    \nabla^i\xi^j\omega_{ij} =- cA^{-4}-6 w^2\frac{\xi^k\mathfrak{Im}(\lambda\d\bar{{\lambda }})_k}{w^4-|\lambda|^4}
\eea

\bea
    \chi_{ijk}(\mathcal{L}_{\xi}\bar{\chi})^{ijk} &= & -6\xi^{l}\Big[\bar{\lambda}\frac{(2w^4+2|\lambda|^4 +11w^2|\lambda|^2)}{(w^2+|\lambda|^2)(w^2-|\lambda|^2)^2}(\d\lambda)_l 
\nonumber \\
&+& \lambda\frac{(7w^4+4|\lambda|^4 + 4w^2|\lambda|^2)}{(w^2+|\lambda|^2)(w^2-|\lambda|^2)^2}(\d\bar{\lambda})_l \nonumber \\
    &+&3w\frac{(2w^2+3|\lambda|^2)}{(w^2-|\lambda|^2)^2}(\d w)_l\Big] -4icA^{-4}
\eea

\bea i\chi_{ijk}(\d\omega)^{ijk} &=& 9\sqrt{2}\xi^l\Big[\bar{\lambda}\frac{(9w^2+|\lambda|^2)}{(w^2-|\lambda|^2)^2}(\d w)_l +w\frac{(w^2+4|\lambda|^2)}{(w^2-|\lambda|^2)^2}(\d\bar{\lambda})_l 
\nonumber \\
&+&5\frac{\bar{\lambda}^2w}{(w^2-|\lambda|^2)^2}(\d\lambda)_l\Big] 
\eea

\bea
    (\d\xi)^{ij}\chi_{ijk} = \frac{\sqrt{2}}{(w^2-|\lambda|^2)}\left[w\frac{\bar{\lambda}}{\lambda} (\d\lambda)_k - w (\d \bar{\lambda})_k \right]
\eea

\bea
    Z = \frac{4}{w^2-|\lambda|^2}\left[w^2 \d\log\rho + w^2 i_{\d\theta} w  -\frac{|\lambda|^2}{w}\d w + \xi\left(\frac{|\lambda|^2}{w} (\d w)_j - w^2 (\d \log\rho)_j \right)\xi^j\right]
\nonumber \\
\eea
   
\bea
    W &=& -\frac{1}{3}\frac{1}{w^2-|\lambda|^2} \bigg[ \frac{1}{w}(w^2 - 4|\lambda|^2)\left( \d w - \xi(\d w)_j\xi^j\right)
\nonumber \\
 &+& (5w^2+4|\lambda|^2)\left(\d\log\rho + i_{\d\theta}w - \xi(\d\log\rho)_j\xi^j\right) \bigg]
\eea

\bea
    \mathcal{L}_{\xi}\xi &=& \frac{1}{3}\bigg[\frac{1}{w}(7w^2 + 2 |\lambda|^2) \left(\d w - \xi  (\d w)_j\xi^j \right) 
\nonumber \\
&+& (w^2 + 2|\lambda|^2) \left( \xi ( \d\log\rho )_j\xi^j - \d\log\rho - i_{\d\theta}w \right)\bigg]
\eea

\bea
    \chi_{ml[i}(\d\xi)^{mn}\bar{\chi}_{j]n}\,^l =\frac{1}{3}cA^{-4} \omega_{ij}\frac{(w^2+|\lambda|^2)}{(w^2-|\lambda|^2)}+2\omega_{ij}\frac{w^2}{(w^4-|\lambda|^4)}\xi^{k}\mathfrak{Im}(\lambda\d\bar{\lambda})_k
\eea

\bea
    8|\lambda|^2(5w^2+|\lambda|^2)\d w + 4w \d\rho^2(w^2+5|\lambda|^2) + 8 w \rho^2\,i_{\d\theta}\omega(w^2-|\lambda|^2) \nonumber \\
 - \xi^i\big[8|\lambda|^2(5w^2+|\lambda|^2)\d w + 4w \d\rho^2(w^2+5|\lambda|^2) + 8 w \rho^2\,i_{\d\theta}\omega(w^2-|\lambda|^2)\big]_i = 0 \ .
\eea

We also obtain a $SU(3)$ invariant expression for the flux $X$. In general, any real
4-form on $M_7$ can be written as
\bea
    X = \t{e}^{\#}\wedge Y + \omega \wedge \sigma + \beta \wedge \chi + \bar{\beta}\wedge\bar{\chi } + X^{\t{TT}}
\eea
where
\begin{itemize}
    \item $\sigma$ is a real two-form;
    \item $\beta$ is a complex one-form, and $\bar{\beta}$ is its complex conjugate;
    \item $Y$ is real 3-form;
    \item $X^{\t{TT}}$ is the traceless (2,2)-part of the flux.
    \end{itemize}

We remark that $X^{\t{TT}}$ is the only part of the flux that is not fixed by the linear system. However, a traceless $(2,2)$ 4-form in 6 dimensions vanishes identically. 
To see this, note that $X^{\t{TT}}$ is dual (in 6 dimensions) to a (1,1) 2-form $R$,
$R = \,*_6 X^{\t{TT}} $. Furthermore, by definition
\bea
    R_{\alpha\bar{\beta}} &=& \frac{1}{4!}\varepsilon_{\alpha\bar{\beta}}\,^{b_1b_2b_3b_4} X^{\t{TT}}_{b_1b_2b_3b_4} = \frac{1}{4}\varepsilon_{\alpha\bar{\beta}}\,^{\mu_1\mu_2\bar{\nu}_1\bar{\nu}_2} X^{\t{TT}}_{\mu_1\mu_2\bar{\nu}_1\bar{\nu}_2} \nonumber \\
     &=& \frac{i}{4}\epsilon_{\alpha}\,^{\bar{\nu}_1\bar{\nu}_2}\varepsilon_{\bar{\beta}}\,^{\mu_1\mu_2} X^{\t{TT}}_{\mu_1\mu_2\bar{\nu}_1\bar{\nu}_2} = \frac{i3!}{4}\delta^{\beta\mu_1\mu_2}_{\alpha\nu_1\nu_2}X^{\t{TT}}_{\mu_1\mu_2}\,^{\nu_1\nu_2} = 0 
\eea
as the contribution from trace terms in the final term vanishes. Hence $R$ vanishes identically, and so $X^{\t{TT}}=0$.

It follows that the flux can be written as
\bea
\label{rel41}
    X = \t{e}^{\#}\wedge Y + \omega \wedge \sigma + \beta \wedge \chi + \bar{\beta}\wedge\bar{\chi } 
\eea
where all of these terms are fixed by the Killing spinor equations. In particular,
the components of the real 2-form $\sigma$  and of the complex 1-form $\beta$ are given by 
\bea
\sigma_{ij} &= &-\frac{(w^2-|\lambda|^2)}{(w^2+|\lambda|^2)}\left(\mathcal{L}_{\xi}\omega   \right)_{ij} \nonumber \\
            &-& 2 \omega_{ij}\frac{\xi^k}{(w^2-|\lambda|^2)(w^2+|\lambda|^2)} \left[ w(w^2+2|\lambda|^2) (\d w)_k  + (2w^2+|\lambda|^2)\mathfrak{Re}(\lambda\d\bar{\lambda})_k     \right]
\nonumber \\
\eea
\bea
\beta_{i} &=&  -3\sqrt{2}\frac{\lambda w}{(w^2-|\lambda|^2)}\left[ \left(\mathcal{L}_{\xi}\xi  \right)_i  +i \left(\mathcal{L}_{\xi}\xi  \right)_j \omega^j \,_i \right]+ \frac{3}{2}\frac{(w^2+|\lambda|^2)}{(w^2-|\lambda|^2)} (\d\xi)^{kj}\bar{\chi}_{kji}  \nonumber   \\ 
&+& i\frac{(w^4+4w^2|\lambda|^2+|\lambda|^4)}{(w^2-|\lambda|^2)(w^2+|\lambda|^2)}\left(\mathcal{L}_{\xi}\omega    \right)^{kj}\bar{\chi}_{kji} \ .
\eea
The real 3-form $Y$ has components
\bea
    Y_{ijk} = q \omega_{[i}\,^l (\d\omega)_{jk]l} + (\omega\wedge V)_{ijk} + h \chi_{ijk} + \bar{h}\bar{\chi}_{ijk}
\eea
where $q$ and $h$ are functions, and $V$ is a real one-form, given by:

\bea
    q = -3\frac{(w^2-|\lambda|^2)}{(w^2+|\lambda|^2)}
\eea
\bea
    V_i = \frac{(w^2-|\lambda|^2)}{(w^2+|\lambda|^2)}\omega_i\,^j(\mathcal{L}_{\xi}\xi)_j
\eea
\bea
    h &=& \frac{3\sqrt{2}}{2}\frac{\xi^k}{(w^2-|\lambda|^2)}\left[\lambda(\d w)_k + \frac{w^3}{(w^2+|\lambda|^2)}(\d\lambda)_k + \frac{w\lambda^2}{(w^2+|\lambda|^2)} (\d\bar{\lambda})_k  \right] 
\nonumber \\
&-& q \frac{i}{6}\bar{\chi}^{ijk}(\d\omega)_{ijk} \ .
\eea

\subsection{SU(3) Invariant Spinor  with $\lambda = 0$, $w \neq 0$}

Next, we consider the special case of the $SU(3)$ invariant spinor
\bea
	\psi_+ = w (1 +\t{e}_{1234})  \qquad w \in \mathbb{R}, \quad w \neq 0 \ .
\eea
The $SU(3)$ covariant geometric conditions obtained from the linear system are:
\bea
    (\d\omega)^{(3,0)} = (\d\omega)^{(0,3)} = 0
\eea
\bea
    \d(w^4\xi) = - \frac{c}{3}A^{-4}\omega w^4
\eea
\bea
    \bar{\chi}^{ijk}\left( \mathcal{L}_{\xi}\omega\right)_{jk} = 0
\eea
\bea
    Z_i = -20 (w^{-1}\d w)_i + 20\xi_i\xi^k(w^{-1}\d w)_k
\eea
\bea
    W_i = 8 (w^{-1}\d w)_i -8\xi_i\xi^k(w^{-1}\d w)_k
\eea
\bea
    \mathfrak{Im}\left(\bar{\chi}^{ijk}(\mathcal{L}_{\xi}\chi)_{ijk} \right) - 4 cA^{-4} = 0
\eea
\bea
    \nabla^i\xi_i = -12 \xi^kw^{-1}(\d w )_k
\eea
\bea
    \nabla^{i}\xi^j\omega_{ij} = -c A^{-4} \ .
\eea
The flux $X$ can be expressed as
\bea
    X = \t{e}^{\#}\wedge Y + \omega \wedge \sigma
\eea
with
\bea
    \sigma = - w^{-2} \mathcal{L}_{\xi} (w^2 \omega)
\eea
and
\bea
    Y_{ijk} = -3 \omega_{[i}\,^l (\d\omega)_{jk]l} + (\omega\wedge V)_{ijk} 
\eea
where
\bea
    V_i  = \omega_i\,^j(\mathcal{L}_{\xi}\xi)_j \ .
\eea

\section{$G_2$ Invariant Spinor }

In this section, we shall consider the case when the stabilizer of $\psi_+$ is $G_2$,
corresponding to the case
\bea
	\psi_+ = w (1 +\t{e}_{1234}) + \lambda\t{e}_{1} + \bar{\lambda}\t{e}_{234} \qquad w \in \mathbb{R}, \ \lambda \in \mathbb{C} \ ,
\eea
with $w^2 = |\lambda|^2$. We shall show that this orbit admits no solutions to the 
Killing spinor equations and the bosonic field equations. To establish this result,
we set $\lambda \equiv e^{i\zeta}w$, where $\zeta$ is a real function. The geometric conditions we obtained by solving the  linear system are:
\bea
    \d w = \d \zeta = 0
\eea
\bea
    \Omega_{\mu,\alpha}\,^{\alpha}=0
\eea
\bea
    \Omega_{\#,\alpha}\,^{\alpha} = i\frac{c}{6}A^{-4}
\eea
\bea
    \Omega_{\#,\alpha\beta}\varepsilon^{\alpha\beta\gamma} = \sqrt{2}\,e^{i\zeta}\,\Omega_{\#,\#}\,^{\gamma}
\eea
\bea
    \Omega_{\bar{\mu},\alpha\beta}\varepsilon^{\alpha\beta\gamma} = \sqrt{2}\,e^{i\zeta}\,\Omega_{\bar{\mu},\#}\,^{\gamma}
\eea
\bea
    2\Omega_{\bar{\mu}}\,^{\gamma\rho} - \sqrt{2}\,e^{-i\zeta}\varepsilon^{\gamma\rho\alpha} \Omega_{\bar{\mu},\#\alpha} + i\sqrt{2}\,\frac{c}{6} e^{-i\zeta}A^{-4}\varepsilon_{\bar{\mu}}\,^{\gamma\rho} =  0 \ .
\eea
Furthermore, we find that all of the components of the flux $X$ vanish,
\bea
    X = 0 \ .
\eea

As $X=0$,  the integrability condition $\Gamma^{j}\left[\nabla_{i},\nabla_j\right]\psi_+$ from \eqref{rel3} implies that
\bea
\label{rel42}
 \Big[\,-\frac{1}{2}\nabla_i\nabla_kA\Gamma^{k} +\frac{c}{2}A^{-4}\nabla_iA\,\tilde{\Gamma}^4  -\frac{c}{12}A^{-4}\nabla_kA\Gamma^{k}\,_i\,\tilde{\Gamma}^4\Big]\psi_+ = 0\ . \
\eea
Multiplying \eqref{rel42} by $\Gamma_l$, we find
\bea
\label{rel47}
 \Big[\,-\frac{1}{2}\nabla_i\nabla_kA(\delta^k_l+ \Gamma_l\,^{k}) +\frac{c}{2}A^{-4}\nabla_iA\Gamma_l\,\tilde{\Gamma}^4  
\nonumber \\
-\frac{c}{12}A^{-4}\nabla_kA(\Gamma_l\,^{k}\,_i +\delta^k_l\Gamma_i-\delta_{li}\Gamma^k )\,\tilde{\Gamma}^4\Big]\psi_+ = 0\ . \
\eea
We next take  the inner product of \eqref{rel47} with $\psi_+$, noting  that the anti-hermitian terms vanish identically as $\psi_+$ is Majorana.
The hermitian part gives 
\bea
\label{rel43}
 \langle\psi_+,\Big[\,-\frac{1}{2}\nabla_i\nabla_lA-\frac{c}{12}A^{-4}\nabla_kA\Gamma_l\,^{k}\,_i \,\tilde{\Gamma}^4\Big]\psi_+\rangle = 0 \ .
\eea
The symmetric part of \eqref{rel43} then gives 
\bea
\label{rel44}
    \nabla_i\nabla_lA ||\psi_+||^2 = 0\quad\Rightarrow\quad  \nabla_i\nabla_lA  = 0 \ ,
\eea
and the antisymmetric part of \eqref{rel43} implies 
\bea
\label{rel45}
    -\frac{c}{12}A^{-4}\nabla_kA\langle\psi_+,\Gamma_{l}\,^k\,_i \psi_+\rangle = 0   
\ .
\eea
The 3-form spinor bilinear in \eqref{rel45} is proportional to the  $G_2$-invariant 3-form $\varphi$ given by
\bea
    \varphi = e^{\#}\wedge\omega - 2i\,\sqrt{2}\, \mathfrak{Im}(e^{i\theta}\chi)\  , 
\eea
\bea
    \varphi_{\#\alpha\bar{\beta}} \equiv -i\delta_{\alpha\bar{\beta}}\qquad\qquad\varphi_{\alpha\beta\gamma}  = -i\sqrt{2}\,e^{i\theta}\varepsilon_{\alpha\beta\gamma} = (\varphi_{\bar{\alpha}\bar{\beta}\bar{\gamma}})^* \ .
\eea
Hence ({\ref{rel45}}) implies that
\bea
\varphi_{lik} \nabla^k A =0
\eea
which in turn implies that $dA=0$, so $A$ is constant. However, from the Einstein field equation \eqref{rel16} we obtain
\bea
\label{rel46}
    3KA^{-1} - \nabla_i\nabla^iA -3A^{-1}\nabla_iA\nabla^iA + \frac{1}{3}c^2A^{-7} + \frac{A}{144}X^2 = 0\ . \
\eea
It is clear that this equation admits no solution in the case for which $A$ is constant and $X=0$, as the LHS is strictly positive. 
Therefore, we conclude that there are no supersymmetric warped product $\t{dS}_4$ solutions for which the spinor $\psi_+$ is $G_2$ invariant.

\section{Conclusion}

We have classified the supersymmetric warped product $\t{dS}_4 \times_w M_7$ solutions in $D=11$ supergravity. To do this, we first integrated explicitly the gravitino equation along the $\t{dS}_4$ directions. This reduces the conditions imposed by supersymmetry to a gravitino-type equation on $M_7$ acting on a Majorana spinor $\psi_+$, whose components depend only on the co-ordinates of $M_7$. Using spinorial geometry techniques, the spinor $\psi_+$ was then simplified to two possible canonical forms by $Spin(7)$ gauge transformations. These two canonical forms have stabilizer subgroups corresponding to $G_2$ and $SU(3)$. In the $G_2$ case, we show that there is no solution to the Killing spinor equations. For the $SU(3)$ case we have determined the 4-form flux in terms of $SU(3)$ invariant geometric structures on $M_7$, as well as determining all of the conditions imposed on the geometry of $M_7$.

This work fully classifies the supersymmetric $\t{dS}_4$ warped product solutions with minimal $N=8$ supersymmetry. It would be interesting to consider the $N=16$ case, as well as the $N=24$ and $N=32$ cases. In particular, for the latter two cases of
$N=24$ and $N=32$ supersymmetry, it is possible to find further conditions on such solutions utilizing the homogeneity theorem analysis constructed in 
\cite{Figueroa-OFarrill:2012kws}. To proceed with this, suppose that there we have $N$ linearly independent solutions $\{ \psi^r : r=1, \dots, N \}$ for $N=24$ or $N=32$ to the gravitino equation
({\ref{rel33b}}). We then consider the integrability condition ({\ref{rel28}}), which implies
\bea
	\left(|\nabla A|^2 - K - \frac{c^2}{9}A^{-6} -\frac{A^2}{(144)^2}\slashed{X}^2 +\frac{2}{3}cA^{-3}\nabla_iA\Gamma^i\tilde{\Gamma}^4 -\frac{1}{18}A\nabla_{i}A\slashed{X}^i\right)\psi^s = 0 \ .
\eea
This implies that
\bea
\langle \psi^r, \Gamma_0 \tilde{\Gamma}^4 \bigg(|\nabla A|^2 - K - \frac{c^2}{9}A^{-6} -\frac{A^2}{(144)^2}\slashed{X}^2 
\nonumber \\
+\frac{2}{3}cA^{-3}\nabla_iA\Gamma^i\tilde{\Gamma}^4 -\frac{1}{18}A\nabla_{i}A\slashed{X}^i\bigg)\psi^s \rangle =0
\eea
and hence
\bea
c \langle \psi^r, \Gamma_0 \Gamma^i \psi^s \rangle \nabla_i A =0 \ .
\eea
On defining vector fields $\Theta^{rs}_i =  \langle \psi^r, \Gamma_0 \Gamma_i \psi^s \rangle$, this implies
\bea
c {\cal{L}}_{\Theta^{rs}} A =0 \ .
\eea
For $N=24$ and $N=32$ solutions, it follows from the homogeneity theorem analysis of
\cite{Figueroa-OFarrill:2012kws} that the $\Theta^{rs}$ span pointwise the tangent space of $M_7$, and hence
\bea
c dA=0 \ .
\eea
If $c \neq 0$, then this implies that $dA=0$. However, ({\ref{rel46}}) implies that
there are no solutions for which $A$ is constant. Hence, for $N=24$ or $N=32$ solutions, we must take $c=0$. 

This determines all possible $N=32$ warped product $\t{dS}_4$ solutions. From
\cite{Figueroa-OFarrill:2002ecq}, where all maximally supersymmetric solutions in
$D=11$ supergravity were determined, the maximally supersymmetric solutions are $\mathbb{R}^{1,10}$ with $F=0$; $\t{AdS}_4 \times {\t{S}}^7$ with 4-form $F$ proportional to the volume form of $\t{AdS}_4$, $\t{AdS}_7 \times {\t{S}}^4$, with 4-form $F$ proportional to the volume form of ${\t{S}}^4$, and a maximally supersymmetric plane wave solution which has 
$F \neq 0$, but $F^2=0$. In terms of possible $N=32$ warped product $\t{dS}_4$ solutions, the condition $c=0$ implies that $F^2 \geq 0$ with equality if and only if $F=0$.
Hence we exclude $\t{AdS}_4 \times {\t{S}}^7$ and the maximally supersymmetric plane wave as $N=32$ warped product $\t{dS}_4$ solutions. It follows that the $N=32$ warped product $\t{dS}_4$ solutions are $\mathbb{R}^{1,10}$ and $\t{AdS}_7 \times {\t{S}}^4$. In particular, it is possible to write both $\mathbb{R}^{1,4}$ and $\t{AdS}_7$ as warped product $\t{dS}_4$ geometries \cite{Gran:2016zxk}. It would be interesting to further understand the possible $N=16$ and $N=24$ warped product $\t{dS}_4$ solutions, though the homogeneity theorem does not apply to the $N=16$ solutions.

\begin{appendices}

\section{Conventions}
\label{conventions}
We use the mostly plus sign signature $\eta = \t{diag}(-,+,\ldots, +)$.  The gamma matrices satisfy 
\bea
	\{\Gamma_A,\Gamma_B\}  = 2g_{AB}\ . \
\eea
	In these conventions, we take
\bea
	\Gamma_{0123456789\#} = \mathbb{I}\  , 
\eea
and consequently the following duality relation holds
\bea
\label{rel7}
	\Gamma_{A_1\ldots A_p} = (-1)^{\frac{(p+1)(p-2)}{2}}\frac{1}{(11-p)!}\varepsilon_{A_1\ldots A_p}\,^{A_{p+1}\ldots A_{11}}\Gamma_{A_{p+1}\ldots A_{11}}\  , 
\eea
where 
\bea
	\varepsilon_{0123456789\#} = +1\ . \ 
\eea
The Hodge star of a $p$-form $\omega$ is defined by
\bea
	*\omega_{A_1\ldots A_{11-p}} = \frac{1}{p!}\varepsilon_{A_1\ldots A_{11-p}}\,^{B_1\ldots B_p}\omega_{B_1\ldots B_p} \ .
\eea
For every $k$-form $\omega$, one can define a Clifford algebra element $\slashed{\omega}$ given by
\bea
	\slashed{\omega} \equiv \omega_{A_1\ldots A_k}\Gamma^{A_1\ldots A_k}\ . \
\eea
In addition, one can define 
\bea
	\slashed{\omega}_C \equiv \omega_{CA_1\ldots A_k}\Gamma^{A_1\ldots A_k}\  , \quad  {\rm and} \ \quad \cancel{\Gamma \omega}_C \equiv \Gamma_{C A_1 \ldots A_k}\omega^{A_1 \ldots A_k}\quad . 
\eea

\section{Spinors from forms}
\label{hSpinorsfromforms}

The Majorana representation  of \emph{Spin}(10,1) can be constructed from the $\emph{Spin}(9,1)$ spinor representations and then adding the tenth gamma matrix $\Gamma_{\#}$. This construction is derived in an explicit representation, in terms of differential forms, in \cite{b8, b10}. We take the space $U$ of 1-forms on $\mathbb{R}^5$, with basis $\{\t{e}_1,\ldots,\t{e}_{5} \}$. The space of Dirac spinors, 
$\Delta_c = \Lambda^*(U \otimes \mathbb{C})$, is identified with the complexified space of multi-forms constructed from this basis. $\Delta_c$ is equipped with a canonical Euclidean Hermitian inner product  $\langle\,~\cdot~,~\cdot~\rangle$

We then take the following representation for the gammma matrices: 
\bea
	\Gamma_0 \eta &=&\, -\t{e}_5 \wedge \eta + i_{\t{e}_5}\eta \quad \Gamma_5 \eta= \, \t{e}_5 \wedge \eta+ i_{\t{e}_5}\eta \nonumber \\ 
	\Gamma_i\eta &= &\, \t{e}_i \wedge \eta + i_{\t{e}_i}\eta \quad i = 1,\ldots,4  \nonumber \\
	\Gamma_{i+5}\eta &= &\, i(\t{e}_i \wedge \eta - i_{\t{e}_i}\eta)\quad
\eea
where $\eta \in \Delta_c$ and $i_{\t{e}_i}$ is the inner derivative along the direction $\t{e}_i$. The tenth gamma matrix can be chosen as 
\bea
\label{gsharp}
	\Gamma_{\#} =  -\Gamma_{0123456789}\ . \
\eea
One can verify that $\Gamma_{\#}^2 = \mathbb{I}$. The gamma matrices satisfy the Clifford Algebra, namely $\Gamma_A\Gamma_B + \Gamma_B\Gamma_A = 2\eta_{AB}\mathbb{I}$.
The Hermitian inner product, acting only on 1-forms, is defined by
\bea
	\langle\,z^a\t{e}_a,w^b\t{e}_b\rangle = (z^a)^*\eta_{ab}w^b\ , \
\eea
and is then extended to the complexified space of multi-forms, $\Delta_c$.

The gamma matrices are chosen such that $\Gamma_0$ is skew-hermitian and $\Gamma_i$, $i = 1 ,\ldots, 9$ are hermitian with respect to $\langle\,~\cdot~,~\cdot~\rangle$. The $Spin(10,1)$ invariant Dirac inner product is defined as
\bea
	D(\eta,\theta) = \langle\,\Gamma_0 \eta, \theta\rangle\ . \
\eea
In eleven dimensions a spinor can be Majorana; the reality condition is 
\bea
\label{spinorsappendix2}
	\eta^* = \Gamma_{6789}\eta \ ,
\eea
where $C = \Gamma_{6789}$ is the charge conjugation matrix, and $C*$ commutes with the gamma matrices, i.e. $C*\Gamma_A = \Gamma_AC*$  \ .
The Dirac representation of  \emph{Spin}(10,1) admits an oscillator basis as 
\bea
	\Gamma_- =&\, \frac{1}{\sqrt{2}}\left(\Gamma_5 - \Gamma_0\right)  = \sqrt{2}\,\t{e}_5\wedge    \quad 
	\Gamma_{\alpha} =  \frac{1}{\sqrt{2}}\left(\Gamma_{\alpha} - i\Gamma_{\alpha + 5} \right)  = \sqrt{2}\,\t{e}_{\alpha} \wedge  \nonumber \\
	\Gamma_+ =&\, \frac{1}{\sqrt{2}}\left(\Gamma_5 + \Gamma_0\right)  = \sqrt{2}\,i_{\t{e}_5}\quad \quad 
	\Gamma_{\bar{\alpha}} =  \frac{1}{\sqrt{2}}\left(\Gamma_{\alpha} + i\Gamma_{\alpha + 5} \right)  = \sqrt{2}\,i_{\t{e}_{\alpha}  }
\eea
and $\Gamma_\sharp$ defined as in ({\ref{gsharp}}). In this oscillator basis, the gamma matrices  satisfy the Clifford Algebra, $\Gamma_A\Gamma_B+ \Gamma_B\Gamma_A = 2\eta_{AB}\mathbb{I}$, with  non-vanishing components are $\eta_{+-} = \eta_{\sharp \sharp}= 1$, $\eta_{\alpha\bar{\beta}} = \delta_{\alpha\bar{\beta}}$.

We note that that $(\Gamma_+)^{\dagger} = \Gamma_-$ and $(\Gamma_{\alpha})^{\dagger} = \Gamma_{\bar{\alpha}}$;
 $(\Gamma^+,\Gamma^{\bar{\alpha}})$ act as creation operators on the Clifford vacuum represented by the 0-degree form 1, where $\Gamma^A = \eta^{AB}\Gamma_B$. A general spinor $\epsilon$ can be written as
\bea
\label{spinorsappendix1}
	\epsilon = \sum_{k=0}^5\frac{1}{k!}\phi_{\bar{a}_1 \ldots\bar{a}_k }\Gamma^{\bar{a}_1 \ldots\bar{a}_k }1 \quad , \quad \bar{a} = +,\bar{\alpha} \ .
\eea

\section{Derivation of equation ({{\ref{rel32}}})}
\label{appendix_constant}
Given the spinor $\Phi$ defined in \eqref{rel37}, we consider ({{\ref{rel32}}}).
In particular we begin by examining the following terms:
\bea
	\nabla_i\Phi + k_1[\t{Eq.}~\eqref{rel3}] \ , 
\eea
where $k_1$, $k_2$ are some constants to be determined. To begin with, note that the terms which are linear in $X$ are:
\bea
\label{rel9}
	&&(1+k_1)\frac{A}{288}\nabla_i\slashed{X} + \frac{1}{576}\nabla_iA \slashed{X} + \frac{(12a+1-4k_1)}{12}\frac{1}{288}cA^{-3}\Gamma_i\slashed{X}\tilde{\Gamma}^4 \nonumber \\
	&+& \frac{(5k_1 - 1-18a)}{432}cA^{-3}\slashed{X}_i\tilde{\Gamma}^4 - \frac{1}{576}\nabla^kA\Gamma_{ki}\slashed{X} +\frac{1}{48}\nabla_kA\Gamma^{kj_1j_2j_3}X_{ij_1j_2j_3} \nonumber \\
	&+& \frac{(3+4k_1)}{48}\nabla^kA\Gamma^{ab}X_{ikab} + \frac{k_1}{72}\nabla_kA\Gamma_{ij_1j_2j_3}X^{kj_1j_2j_3}\ . \
\eea
In order to set to zero the term involving $\nabla_i\slashed{X}$, we set $k_1=-1$.
Having done so, we then consider imposing the condition
\bea
\label{6666rel32}
	\nabla_i\Phi - [\t{Eq.}~\eqref{rel3}] + k_2A^{-1}\Gamma^i[\t{Eq.}~\eqref{rel5}] 
	 + q_1 \cancel{\Gamma X}_i\Phi + q_2\slashed{X}_i\Phi
\nonumber \\
 + q_3cA^{-4}\Gamma_i\tilde{\Gamma}^4\Phi + q_4A^{-1}\nabla_kA\Gamma_i\Gamma^k\Phi + q_5 A^{-1}\nabla_iA\Phi = 0 \ ,
\eea 
and compute all of the terms on the LHS, choosing the constants $a, k_2, q_1, q_2, q_3, q_4, q_5$ so that the identity above holds.
The terms involving the quadratic contribution of $X$ are
\bea
	&&A\Big[ \frac{(4k_2 +1-288q_1)}{1152} \Gamma^{l_1l_2}\,_{ij_1j_2}X_{j_3j_4l_1l_2}X^{j_1j_2j_3j_4} 
\nonumber \\
&+& \frac{(  288q_1 - 72q_2-3 )}{1728}\Gamma_{j_1j_2j_3}\,^{l_1l_2}X_{ij_4l_1l_2}X^{j_1j_2j_3j_4}  \nonumber\\
	&+&\frac{(8k_2-288q_1-72q_2-1)}{576}\Gamma^{abm}X_{impq}X^{pq}\,_{ab}-\frac{(1+4k_2-288q_1)}{12}\frac{1}{288}\Gamma_iX^2  \nonumber\\
	&+& \frac{(3-288q_1+ 72q_2)}{864}\Gamma_{j_1}X_{ij_2j_3j_4}X^{j_1j_2j_3j_4} \Big] \ . \nonumber\\
\eea
The terms involving the linear contribution of $X$ are
\bea
	 &&\frac{(1-288q_1+2q_4 +2q_5)}{576}\nabla_iA \slashed{X} + \frac{(12a+5+12q_3 +3456aq_1)}{12}\frac{1}{288}cA^{-3}\Gamma_i\slashed{X}\tilde{\Gamma}^4  \nonumber\\
	&+&  \frac{(288q_1-2q_4-1)}{576}\nabla^kA\Gamma_{ki}\slashed{X} +\frac{(1+24q_2-96q_1)}{48}\nabla_kA\Gamma^{kj_1j_2j_3}X_{ij_1j_2j_3} \nonumber\\
	&+&\,\frac{(8k_2-1-288q_1-72q_2)}{48}\nabla^kA\Gamma^{ab}X_{ikab} - \frac{(1+4k_2-288q_1)}{72}\nabla_kA\Gamma_{ij_1j_2j_3}X^{kj_1j_2j_3}\nonumber\\
	&-& \frac{(6+18a+1728aq_1 -432q_2a)}{432}cA^{-3}\slashed{X}_i\tilde{\Gamma}^4\quad.
\eea

The terms involving no contribution of $X$ are
\bea
	&& {1 \over 24}(1- 16k_2-24aq_4+ 12q_3)cA^{-4}\nabla^kA\Gamma_k\Gamma_i\tilde{\Gamma}^4 
\nonumber \\
&+& {1 \over 12} (16k_2-7-36a-12   q_3 +24aq_4+12aq_5)cA^{-4}\nabla_iA\tilde{\Gamma}^4 + 
\nonumber\\
	&+& (k_2-\frac{1}{2}q_4)A^{-1}|\nabla A|^2\Gamma_i -k_2A^{-1} K\Gamma_i -\frac{1}{2}q_5A^{-1}\nabla_kA\nabla_iA\Gamma^k
\nonumber \\
&-&{1 \over 36}(3a+4k_2+36q_3a)c^2A^{-7}\Gamma_i \ .
\nonumber \\
\eea
By requiring that all terms in the above expressions should vanish, we are able to determine the constant values, that are 
\bea
	a  = -\frac{1}{6} \quad q_1  = \frac{1}{288} \quad q_2  = -\frac{1}{36}  \quad q_3  = -\frac{1}{12}\quad k_2  =q_4= q_5 = 0 \ .
\eea

\section{KSE Linear System - SU(3) Stabilizer}
\label{linearsystem}
The linear system associated to the KSEs~\eqref{rel33b}, with the spinor given by
\bea
	\psi_+ = w (1 +\t{e}_{1234}) + \lambda\t{e}_{1} + \bar{\lambda}\t{e}_{234} \qquad w \in \mathbb{R}, \ \lambda \in \mathbb{C} 
\eea
is as follows:
\bea 
\label{grel1}
\partial_{\#}w + \frac{w}{2}\Omega_{\#,\alpha}\,^{\alpha} + \frac{1}{24}w X_{\alpha}\,^{\alpha}\,_{\beta}\,^{\beta} -i \frac{c}{12}A^{-4}w-\frac{\sqrt{2}}{3}\bar{\lambda}X_{\# 234} = 0
\eea
 \bea
 \partial_{\#}\lambda +\frac{\lambda}{2}\Omega_{\#,\alpha}\,^{\alpha}-\frac{1}{24}\lambda X_{\alpha}\,^{\alpha}\,_{\beta}\,^{\beta} + \frac{\sqrt{2}}{3}wX_{\# 234} - i \frac{c}{12}A^{-4}\lambda = 0
  \eea
\bea
  w\Omega_{\#,\alpha\beta}\varepsilon^{\gamma\alpha\beta} -\sqrt{2}\lambda\Omega_{\# ,\#}\,^{\gamma} +\frac{\sqrt{2}}{3}\lambda X_{\# \alpha}\,^{\alpha\gamma} +\frac{w}{3}X^{\gamma}\,_{234}=0 
  \eea
   \bea
  \bar{\lambda}\varepsilon^{\gamma\alpha\beta} \Omega_{\#\alpha\beta} - \sqrt{2}w\Omega_{\#,\#}\,^{\gamma} -\frac{w\sqrt{2}}{3}X_{\#\alpha}\,^{\alpha\gamma}  - \frac{\bar{\lambda}}{3}X^{\gamma}\,_{234} = 0
  \eea
  \bea
  \partial_{\mu}w + \frac{1}{2}w\Omega_{\mu,\alpha}\,^{\alpha} - \frac{1}{4}wX_{\mu\#\alpha}\,^{\alpha} = 0 
  \eea
 \bea
  \partial_{\mu}\lambda + \frac{1}{2}\lambda\Omega_{\mu,\alpha}\,^{\alpha} + \frac{1}{4}\lambda X_{\mu\#\alpha}\,^{\alpha} = 0 
  \eea
  \bea
  \partial_{\mu}w -\frac{w}{2}\Omega_{\mu,\alpha}\,^{\alpha} - \frac{w}{12}X_{\#\mu\alpha}\,^{\alpha} - \frac{\sqrt{2}}{3}\lambda X_{\mu\bar{2}\bar{3}\bar{4}} = 0
  \eea
  \bea
 \partial_{\mu}\bar{\lambda} - \frac{\bar{\lambda}}{2}\Omega_{\mu,\alpha}\,^{\alpha} + \frac{1}{12}\bar{\lambda}X_{\#\mu\alpha}\,^{\alpha} +\frac{\sqrt{2}}{3}wX_{\mu\bar{2}\bar{3}\bar{4}} = 0
  \eea
  
\bea
w\Omega_{\bar{\mu},\alpha\beta}\varepsilon^{\alpha\beta\gamma} - \sqrt{2}\lambda\Omega_{\bar{\mu},\#}\,^{\gamma} - \frac{w}{2}\varepsilon^{\gamma\alpha\beta}X_{\bar{\mu}\#\alpha\beta} -\frac{w}{3}\varepsilon_{\bar{\mu}}\,^{\gamma\rho}X_{\#\rho\alpha}\,^{\alpha} +\frac{\sqrt{2}}{6}\lambda \varepsilon_{\bar{\mu}}\,^{\gamma\rho}X_{\rho\bar{2}\bar{3}\bar{4}} = 0 
\nonumber \\
 \eea
 \bea
 w\Omega_{\bar{\mu},}\,^{\gamma\rho} -\frac{\sqrt{2}}{2}\bar{\lambda}\varepsilon^{\gamma\rho\alpha}\Omega_{\bar{\mu},\#\alpha} + \varepsilon_{\bar{\mu}}\,^{\gamma\rho}\Big[ \frac{w}{6}X_{\#\bar{2}\bar{3}\bar{4}}- \frac{\bar{\lambda}\sqrt{2}}{24}X_{\alpha}\,^{\alpha}\,_{\beta}\,^{\beta}
\nonumber \\
 + i\frac{c\sqrt{2}}{12}\bar{\lambda}A^{-4}\Big]  - \frac{\sqrt{2}}{4}\bar{\lambda}\varepsilon^{\gamma\rho\alpha}X_{\bar{\mu}\alpha\beta}\,^{\beta} = 0
 \eea
 
\bea
 \bar{\lambda}\Omega_{\bar{\mu},\alpha\beta}\varepsilon^{\alpha\beta\gamma} - \sqrt{2}w\Omega_{\bar{\mu},\#}\,^{\gamma} + \frac{\bar{\lambda}}{2}\varepsilon^{\gamma\alpha\beta}X_{\bar{\mu}\#\alpha\beta} + \frac{\bar{\lambda}}{3}\varepsilon_{\bar{\mu}}\,^{\gamma\rho}X_{\#\rho\alpha}\,^{\alpha} -\frac{\sqrt{2}}{6}w\varepsilon_{\bar{\mu}}\,^{\gamma\rho}X_{\rho\bar{2}\bar{3}\bar{4}} = 0 
\nonumber \\
\eea
\bea
 \label{grel2}
 \lambda\Omega_{\bar{\mu},}\,^{\gamma\rho} -\frac{\sqrt{2}}{2}w\varepsilon^{\gamma\rho\alpha}\Omega_{\bar{\mu},\#\alpha} + \varepsilon_{\bar{\mu}}\,^{\gamma\rho}\Big[ \frac{\sqrt{2}w}{24}X_{\alpha}\,^{\alpha}\,_{\beta}\,^{\beta} 
\nonumber \\
+ i\frac{c\sqrt{2}}{12}wA^{-4} -\frac{\lambda}{6}X_{\#\bar{2}\bar{3}\bar{4}}\Big]  + \frac{\sqrt{2}}{4}w\varepsilon^{\gamma\rho\alpha}X_{\bar{\mu}\alpha\beta}\,^{\beta} = 0 \ .
 \eea

\subsection{Solution for $\lambda \neq 0$, $w \neq 0$}

 From the linear system ({\ref{grel1}})-({\ref{grel2}}), we find that the components of the flux are given by the following expressions 
\bea
X_{\#\alpha}\,^{\alpha\gamma} = \frac{3}{w^2 -|\lambda|^2}\left[\sqrt{2}w\bar{\lambda} \varepsilon^{\gamma\alpha\beta}\Omega_{\#,\alpha\beta} -(w^2 +|\lambda|^2)\Omega_{\#,\#}\,^{\gamma} \right]
\eea
\bea
    X_{\bar{\gamma}234} = \frac{3}{w^2 -|\lambda|^2}\left[2\sqrt{2}\lambda w \Omega_{\#,\#}\,^{\gamma}-(w^2 +|\lambda|^2) \Omega_{\#,\alpha\beta} \varepsilon^{\gamma\alpha\beta} \right]
\eea

\bea
    X_{\#234 } = \frac{1}{w^2 -|\lambda|^2}\left[-\frac{3}{\sqrt{2}}\lambda \partial_{\#}w  -\frac{3}{\sqrt{2}}w\partial_{\#}\lambda -\frac{3}{\sqrt{2}}w\lambda\Omega_{\#,\alpha}\,^{\alpha} + i\frac{c}{2\sqrt{2}}A^{-4}\lambda w  \right]
\eea
\bea
    X_{\alpha}\,^{\alpha}\,_{\beta}\,^{\beta} &=&  \frac{1}{w^2 -|\lambda|^2}\big[ -24w\partial_{\#}w -24 \bar{\lambda}\partial_{\#}\lambda 
\nonumber \\
 &-& 12 (w^2 +|\lambda|^2)\Omega_{\#,\alpha}\,^{\alpha} +2icA^{-4}    (w^2 +|\lambda|^2)  \big]
\eea

\bea
    X_{\bar{\mu}\#\alpha\beta} = 2\Omega_{\bar{\mu},\alpha\beta} -\sqrt{2}\frac{\lambda}{w}\varepsilon_{\gamma\alpha\beta}\Omega_{\bar{\mu},\#}\,^{\gamma} + \delta_{\bar{\mu}[\alpha} \left(2\Omega_{\#,\#\beta]} - \sqrt{2}\frac{\lambda}{w}\varepsilon_{\beta]\delta\sigma}\Omega_{\#}\,^{\delta\sigma}  \right)
\eea

\bea
 X_{\bar{\mu}\alpha\beta}\,^{\beta} &=& \sqrt{2}\frac{w}{\bar{\lambda}}\varepsilon_{\alpha\gamma\rho}\Omega_{\bar{\mu},}\,^{\gamma\rho} - 2\Omega_{\bar{\mu},\#\alpha} + \frac{\delta_{\bar{\mu}\alpha}}{w^2 - |\lambda|^2}\bigg( 3w\partial_{\#}w - \frac{w^2}{\bar{\lambda}}\partial_{\#}\bar{\lambda}
\nonumber \\
 &+& 4\bar{\lambda}\partial_{\#}\lambda + \Omega_{\#,\alpha}\,^{\alpha} (3w^2 + 2|\lambda|^2) - i\frac{c}{6}A^{-4}(4|\lambda|^2 + w^2)  \bigg) \ .
\eea
From the linear system ({\ref{grel1}})-({\ref{grel2}}), we also find the following geometric conditions:
\bea
&&4(w^2-|\lambda|^2)\partial_{\mu}w + 2w(w^2-|\lambda|^2)\Omega_{\mu,\alpha}\,^{\alpha} \nonumber\\ &-& 3\sqrt{2}w^2\lambda\varepsilon_{\mu\alpha\beta}\Omega_{\#,}\,^{\alpha\beta} + 3w(w^2+|\lambda|^2)\Omega_{\#,\#\mu} = 0
\eea
 \bea
&&4(w^2-|\lambda|^2)\partial_{\mu}\lambda + 2\lambda(w^2-|\lambda|^2)\Omega_{\mu,\alpha}\,^{\alpha} \nonumber\\&+& 3\sqrt{2}w\lambda^2\varepsilon_{\mu\alpha\beta}\Omega_{\#,}\,^{\alpha\beta} - 3\lambda(w^2+|\lambda|^2)\Omega_{\#,\#\mu} = 0
\eea
\bea
    &&4(w^2-|\lambda|^2)\partial_{\mu}w - 2w(w^2-|\lambda|^2)\Omega_{\mu,\alpha}\,^{\alpha} \nonumber\\&+&\sqrt{2}\lambda(5w^2+4|\lambda|^2)\varepsilon_{\mu\alpha\beta}\Omega_{\#,}\,^{\alpha\beta} - w(w^2+17|\lambda|^2)\Omega_{\#,\#\mu} = 0
\eea
\bea
    &&4(w^2-|\lambda|^2)\partial_{\mu}\bar{\lambda} - 2\bar{\lambda}(w^2-|\lambda|^2)\Omega_{\mu,\alpha}\,^{\alpha} \nonumber\\&-&\sqrt{2}w(4w^2+5|\lambda|^2)\varepsilon_{\mu\alpha\beta}\Omega_{\#,}\,^{\alpha\beta} + \bar{\lambda}(17w^2+|\lambda|^2)\Omega_{\#,\#\mu} = 0
\eea

\bea
    &&2\bar{\lambda}w\varepsilon^{\gamma\alpha\beta}\Omega_{\bar{\mu},\alpha\beta} -\sqrt{2}(w^2+|\lambda|^2)\Omega_{\bar{\mu},\#}\,^{\gamma} \nonumber\\&-&2\bar{\lambda}w\varepsilon^{\gamma\beta}\,_{\bar{\mu}}\Omega_{\#,\#\beta} + \sqrt{2}(w^2+|\lambda|^2)\Omega_{\#,\bar{\mu}}\,^{\gamma} = 0
\eea

  \bea
  &&8(|\lambda|^2 +w^2)\Omega_{\bar{\mu}}\,^{\gamma\rho} -8\sqrt{2}w\bar{\lambda}\varepsilon^{\gamma\rho\alpha}\Omega_{\bar{\mu},\#\alpha} \nonumber\\&+& \sqrt{2}\varepsilon_{\bar{\mu}}\,^{\gamma\rho}\Big[2w\bar{\lambda}\Omega_{\#,\alpha}\,^{\alpha} - 2 w\partial_{\#}\bar{\lambda} - 2\bar{\lambda}\partial_{\#}w  + icA^{-4}\bar{\lambda}w\Big] = 0
  \eea

\bea
    6(\bar{\lambda}\partial_{\#}\lambda - \lambda\partial_{\#}\bar{\lambda} )+ 6(w^2+|\lambda|^2)\Omega_{\#,\alpha}\,^{\alpha} - icA^{-4}(w^2+|\lambda|^2) = 0
\eea

\bea
    &&2w(w^2-|\lambda|^2)\Omega^{\alpha}\,_{,\alpha\gamma} + \sqrt{2}\lambda(w^2-|\lambda|^2)\Omega^{\alpha}\,_{,\#}\,^{\beta}\varepsilon_{\alpha\beta\gamma} \nonumber\\
    &-& w\Omega_{\#,\#\gamma}(w^2+5|\lambda|^2) + \sqrt{2}\lambda(2w^2+|\lambda|^2)\varepsilon_{\gamma\alpha\beta}\Omega_{\#,}\,^{\alpha\beta} = 0
\eea
\bea
  && 2\sqrt{2}w (w^2-|\lambda|^2)\Omega^{\alpha\beta\gamma}\varepsilon_{\alpha\beta\gamma} - 4\bar{\lambda}(w^2-|\lambda|^2)\Omega^{\alpha}\,_{,\#\alpha}-30w\bar{\lambda}\partial_{\#}w \nonumber\\
   &-& 6w^2\partial_{\#}\bar{\lambda} -24\bar{\lambda}^2\partial_{\#}\lambda -6 \bar{\lambda}(w^2+2|\lambda|^2)\Omega_{\#,\alpha}\,^{\alpha} + 3icA^{-4}\bar{\lambda}w^2 =  0
\eea
\bea
     &&(w^2-|\lambda|^2)^2\left(\Omega_{\mu,\#\bar{\alpha}} - \Omega_{\bar{\alpha},\#\mu}\right)+ \delta_{\mu\bar{\alpha}} \big[ (3w^2+2|\lambda|^2)(\lambda\partial_{\#}\bar{\lambda} - \bar{\lambda}\partial_{\#}\lambda) \nonumber\\&-& 2 (w^4 + 3w^2|\lambda|^2 + |\lambda|^4)\Omega_{\#,\alpha}\,^{\alpha} + i \frac{c}{3}A^{-4}(2w^4 + w^2|\lambda|^2+2|\lambda|^4) \big] =0 \ . \nonumber \\
\eea
\subsection{Solution for $\lambda = 0$, $w \neq 0$}
We next present the components of the flux and the geometric conditions associated to the KSEs~\eqref{rel33b}, with the spinor given by
\bea
	\psi_+ = w (1 +\t{e}_{1234})  \qquad w \in \mathbb{R}\ . 
\eea 
We find that the components of the flux are given by the following expressions 
\bea
    X_{\#234} = X_{\bar{\mu}234} = 0
\eea
\bea
    X_{\#\alpha}\,^{\alpha\gamma} = -3 \Omega_{\#,\#}\,^{\gamma}
\eea
\bea
    X_{\alpha}\,^{\alpha}\,_{\beta}\,^{\beta} = -24w^{-1}\partial_{\#}w
\eea
\bea
    X_{\bar{\mu}\#\alpha\beta} = 2 \Omega_{\bar{\mu},\alpha\beta} + 2 \delta_{\bar{\mu}[\alpha} \Omega_{\#,\#\beta]}
\eea
\bea
    X_{\bar{\mu}\nu\beta}\,^{\beta} =  \left(\Omega_{\bar{\mu},\#\nu} + \Omega_{\nu,\#\bar{\mu}}\right) +  4 \delta_{\bar{\mu}\nu}w^{-1}\partial_{\#}w \ .
\eea
Furthermore, we find that the geometric conditions are given  by the following expressions 
\bea
    \Omega_{\#,\alpha\beta}\varepsilon^{\gamma\alpha\beta} = 0
\eea
\bea
    \Omega_{\#,\#\mu} = -4w^{-1}\partial_{\mu}w
\eea
\bea
    \Omega_{\mu,\alpha}\,^{\alpha} = 4w^{-1}\partial_{\mu}w
\eea
\bea
    \Omega_{\mu,\alpha\beta} = 0
\eea
\bea
    \Omega_{\mu,\#\beta} = 0
\eea
\bea
    \Omega^{\alpha}\,_{\alpha\beta} = -2 \frac{\partial_{\beta}w}{w}
\eea
\bea
    \Omega^{\alpha}\,_{\#\alpha} - 6 \frac{\partial_{\#}w}{w} - \frac{i}{2}cA^{-4} = 0
\eea
\bea
    \left(\Omega_{\bar{\alpha},\#\beta}-\Omega_{\beta,\#\bar{\alpha}}\right) - i\frac{c}{3}A^{-4}\delta_{\bar{\alpha}\beta} = 0
\eea
\bea
    \Omega_{\#,\alpha}\,^{\alpha} = i\frac{c}{6}A^{-4} \ .
\eea
\section{Covariant relations}
\label{covariantexpression}
In this Appendix, we present the main relations used to covariantize the linear system.
These expressions relate spin connection terms to SU(3)-covariant terms
involving the the $SU(3)$ invariant 1-forms $\xi$, $\omega$ and $\chi$,
their Lie derivatives with respect to $\xi$, and also the Lee forms $W$ and $Z$: 
\bea
    \nabla_{\#}\xi_{\mu}  =(\mathcal{L}_{\xi}\xi)_{\mu} = - \Omega_{\#,\#\mu}
\eea
\bea
    \nabla_{\bar{\alpha}}\xi_{\beta}  = - \Omega_{\bar{\alpha},\#\beta}
\eea
\bea
    \nabla_{\#}\omega_{\alpha\beta} = 2i\Omega_{\#,\alpha\beta}
\eea
\bea
    W_{\alpha}  =- \Omega_{\#,\#\alpha} - 2 \Omega^{\beta}\,_{\beta\alpha}
\eea
\bea
    W_{\bar{\alpha}}  = -\Omega_{\#,\#\bar{\alpha}} - 2 \Omega^{\bar{\beta}}\,_{\bar{\beta}\bar{\alpha}}
\eea
\bea
    Z_{\bar{\rho}} = 2 \Omega_{\#,\#\bar{\rho}} + 2 \Omega_{\bar{\rho},\gamma}\,^{\gamma} + 2 \Omega^{\bar{\gamma}}\,_{\bar{\gamma}\bar{\rho}} 
\eea
\bea
      Z_{\rho} = 2 \Omega_{\#,\#\rho} + 2 \Omega_{\rho,\bar{\gamma}}\,^{\bar{\gamma}} + 2 \Omega^{\gamma}\,_{\gamma\rho} 
\eea
\bea
    (\mathcal{L}_{\xi}\omega) _{\alpha\beta} = 2i\Omega_{\#,\alpha\beta} + i(\d \xi)_{\alpha\beta}
\eea
\bea
    (\mathcal{L}_{\xi}\omega) _{\bar{\alpha}\bar{\beta}} = -2i\Omega_{\#,\bar{\alpha}\bar{\beta}} - i(\d \xi)_{\bar{\alpha}\bar{\beta}}
\eea
\bea
    (\mathcal{L}_{\xi}\omega)_{\alpha\bar{\beta}} = i(\Omega_{\bar{\beta},\#\alpha} + \Omega_{\alpha,\#\bar{\beta}})
\eea
\bea
    (\mathcal{L}_{\xi}\chi)_{\alpha\beta\gamma} =3\Omega_{\#,[\gamma}\,^{\lambda}\varepsilon_{\alpha\beta]\lambda}-3\Omega_{[\gamma,\#}\,^{\lambda}\varepsilon_{\alpha\beta]\lambda} = (\Omega_{\#,\lambda}\,^{\lambda} - \Omega_{\lambda,\#}\,^{\lambda})\varepsilon_{\alpha\beta\gamma}
\eea
\bea
       (\mathcal{L}_{\xi}\chi)_{\alpha\beta\bar{\gamma}} = (\Omega_{\#,\bar{\gamma}}\,^{\lambda} - \Omega_{\bar{\gamma},\#}\,^{\lambda})\varepsilon_{\lambda\alpha\beta}
\eea
\bea
\mathcal{L}_{\xi}\bar{\chi} = (\mathcal{L}_{\xi}\chi )^* \ .
\eea
The spin-connection components are rewritten in terms of those covariant quantities as
\bea
    \Omega_{\mu,\alpha}\,^{\alpha} = -\frac{1}{2}\left((\mathcal{L}_{\xi}\xi)_{\mu}  +W_{\mu}+Z_{\mu}\right) 
\eea
\bea
    \Omega_{\#,\#\mu} = -\nabla_{\#}\xi_{\mu} = -i\nabla_{\#}\omega_{\#\mu} = -(\mathcal{L}_{\xi}\xi)_{\mu}
\eea
\bea
    \Omega_{\#,\alpha\beta} = -\frac{i}{2}\nabla_{\#}\omega_{\alpha\beta} = -\frac{1}{2}\left(i(\mathcal{L}_{\xi}\omega)_{\alpha\beta} + (\d\xi)_{\alpha\beta} \right)
\eea
\bea
    \Omega_{\bar{\alpha},\beta\gamma} = -\frac{i}{2}(\d\omega)_{\bar{\alpha}\beta\gamma
    }
\eea
\bea
    \Omega_{\bar{\alpha},\#\bar{\beta}} = \frac{1}{2}\left(i (\mathcal{L}_{\xi}\omega)_{\bar{\alpha}\bar{\beta}} -(\d\xi)_{\bar{\alpha}\bar{\beta}} - \varepsilon_{\bar{\beta}}\,^{\gamma\rho}(\mathcal{L}_{\xi}\chi)_{\gamma\rho\bar{\alpha}}\right)
\eea
\bea
    \Omega_{\bar{\alpha}\bar{\beta}\bar{\gamma}} = \frac{i}{2}\nabla_{\bar{\alpha}}\omega_{\bar{\beta}\bar{\gamma}}
\eea
\bea
    \varepsilon_{\alpha\beta\gamma} \nabla^{\alpha}\omega^{\beta\gamma} = \frac{1}{3} \varepsilon_{\alpha\beta\gamma} (\d\omega)^{\alpha\beta\gamma} 
\eea
\bea
    \Omega_{\bar{\alpha},\#\beta} = -\nabla_{\bar{\alpha}}\xi_{\beta} = -i\nabla_{\bar{\alpha}}\omega_{\#\beta}
\eea
\bea
    \Omega_{\#,\rho}\,^{\rho}=-\frac{1}{6}\varepsilon_{\alpha\beta\gamma}(\mathcal{L}_{\xi}\bar{\chi})^{\alpha\beta\gamma} + \nabla^{\lambda}\xi_{\lambda}
\eea
\bea
    \Omega^{\beta}\,_{\beta\alpha} = \frac{1}{2}\left((\mathcal{L}_{\xi}\xi)_{\alpha} - W_{\alpha}\right)
\eea
\bea
    \Omega^{[\alpha}\,_{\#}\,^{\beta]} = -\frac{1}{2}(\d\xi)^{\alpha\beta}
\eea
\bea
   \Omega^{\alpha}\,_{\#\alpha} = - \nabla^{\alpha}\xi_{\alpha}  = -i\nabla^{\alpha}\omega_{\#\alpha} \ .
\eea

\end{appendices}

\section*{Acknowledgments}

DF is partially supported by the STFC DTP Grant ST/S505742.

\section*{Data Management}

No additional research data beyond the data presented and cited in this work are needed to validate the research findings in this work.

\end{document}